\documentclass[%
 aip,
 amsmath,amssymb,
 reprint,%
]{revtex4-1}

\usepackage{graphicx}
\usepackage{dcolumn}
\usepackage{bm}

\usepackage[utf8]{inputenc}
\usepackage[T1]{fontenc}
\usepackage{mathptmx}

\begin{document}

\preprint{AIP/123-QED}

\title[Cavitation and ventilation modalities during ditching]{Cavitation and 
ventilation modalities during ditching}

\author{A. Iafrati}
 \email{alessandro.iafrati@cnr.it.}
 \affiliation{National Research Council - Institute of Marine
Engineering - Rome, Italy.}
\author{S. Grizzi}%
\affiliation{ 
National Research Council - Institute of Marine
Engineering - Rome, Italy.
}%


\date{\today}

\begin{abstract}
The flow taking place in the rear part of the fuselage during the
emergency landing on water is investigated experimentally in realistic
conditions.
To this aim, tests on a double curvature specimen have been performed at
horizontal velocities ranging from 21 m/s to 45 m/s.
Tests data highlight different cavitation and/or ventilation
modalities which are highly dependent to the horizontal velocity, with
substantial variations in the flow features occurring with velocity
variations of few meter per second.
For the specimen considered here, the inception of the cavitation is found
at about 30 m/s, confirming that scaled model tests performed at small
horizontal velocities are unable to capture the hydrodynamics correctly.
By comparing pressure data, underwater movies and force measurements, it
is shown that the transition from cavitation to ventilation condition has
a significant effect of the longitudinal distribution of the loading which,
together with inertia, aerodynamic loads and engine thrust, governs the
aircraft dynamics.
\end{abstract}

\maketitle


\section{\label{sec:intro}Introduction\protect}

The ditching is an emergency procedure consisting in a controlled landing
on water. The event is well known and the risk of ditching is
reminded on every flight before the take off.
Whereas according to the statistics the event is fortunately rare,
the ditching became popular after the Hudson river event in February 2009.

Of course for a successful event it is important that all influencing
factors are well settled. One is certainly the pilot skill, but, equally
important, the aircraft has to be properly designed.
The design for ditching is not aimed at preventing failures in the
structure
but mainly at ensuring a floatation time long enough to allow a safe
evacuation of the plane.
The aircraft design cannot be validated by full scale tests as they would
be too expensive and impractical. Nowadays, the tendency is to use
computational tools \cite{siemann2017, siemann2018,qu2015,bisagni2018}.
However, the hydrodynamics of ditching and the resulting interaction with
the structure, are highly nonlinear and challenging even for the most
sophisticated numerical models available at this time and a careful
validation versus representative experiments is deemed necessary.

Owing to the large dimensions and velocities involved, laboratory
experiments are
often conducted on scaled models by exploiting the Froude similarity that
guarantees the correct reproduction of the ratio between
inertia and gravitational forces.
With such a scaling, the test velocity is reduced as the square
root of the length. Of course viscous and surface tension effects, which
are accounted for by the Reynolds and Weber similarities, are not properly
scaled\cite{landau1959}, but their effects are minor, at least in the early
stage of impact\cite{Moghisi1981}.
Much more relevant are instead the effects of cavitation and ventilation
which are not properly reproduced because of the different Euler number.
Their relevance in the ditching phase was already discussed in previous
studies\cite{smith1957,zhang2012}.
Finally, when large structural deformations occur, the fluid loading
is significantly different from that occurring on a rigid component and
a proper
scaling of the structure that enables a correspongly scaled hydrodynamic
loading is impossible to be achieved\cite{climent2006}.

A unique opportunity to perform highly representative tests is offered by
the High Speed Ditching Facility (HSDF), available at CNR-INM, where
realistic specimen, albeit of limited size, may be tested at nearly full
scale velocity\cite{iafratijfs2015}.
Thanks to the maximum achievable speed of 47 m/s, which is in the range of
the ditching velocity for small passengers or cargo aircrafts, it
allows reproducing all physical phenomena without any scaling issue.

In Ref. \onlinecite{iafratijfm2016} the water impact at high velocity of a thick
rectangular plate is investigated.
The high pressure peak at the spray root and the correspondingly large
induced loads are analyzed and compared to theoretical solutions,
showing the important role played by the three-dimensional effects on
the of pressures and loads.
The relevance of fluid-structure interaction is clearly highlighted in
Ref. \onlinecite{iafratinav2015} where ditching tests on flat plates 
undergoing very
large and even permanent out-of-plane deformations are reported.
The effects of the transverse curvature are analyzed instead
in Ref. \onlinecite{iafrati2018}.
It is shown that convex curvatures, enhancing the possibility for the
fluid to escape from the side, lead to a reduction of pressure and loads
compared to the flat plate. In the case of a concave plate, the pressures
peaks in the middle are much lower that in the flat plate case, but the
total loading is essentially the same.

For the aircraft dynamics the longitudinal curvature is far more important.
The effect of the fuselage shape on the aircraft dynamics was already
investigated in Ref. \onlinecite{mcbride1953}. However, the tests, 
being performed at
model scale, do not provide an accurate reproduction of the cavitation and
ventilation phenomena that might take place at the rear.
The occurrence of suction forces at the rear of the fuselage induced by the
shrinking of the shape and by the longitudinal curvature is discussed
in Ref. \onlinecite{tassin2013}, albeit based on a simplified representation 
of the problem.

With the aim of investigating the flow taking place in the rear part of the
fuselage, tests a double curvature specimen are performed.
Results are provided in terms of pressures and loads, supported by
underwater
visualizations, and clearly shown the occurrence of cavitation/ventilation
phenomena, although different scenarios are found depending on the test
speed.
It is found that the change from one condition to another is rather sharp
and happens within velocity variations of few m/s.

Although the present study is primarily motivated by the aircraft ditching,
the observed cavitation/ventilation modalities are of great interest in
other
applications, notably high-speed, supercavitating, underwater vehicles
\cite{wang2017} or drag reduction through air lubrication
\cite{ceccio2010} where the
occurrence of cavitation/ventilation conditions, if properly controlled,
substantially enhances the vehicles' efficiency.

\section{\label{sec:expsetup}Experimental setup and instrumentation\protect}

Experimental tests were performed at the High Speed Ditching Facility: it
is composed by a guide suspended over a water basin. The trolley bringing
the specimen to be tested runs along the guide.
The guide can be rotated enabling different vertical to horizontal velocity
ratios.
The specimen to be tested and the acquisition box are connected
to the trolley. The trolley is accelerated by a set of elastic cords and,
shortly
before the contact with water, it is left to run freely thus ensuring that
no external forces are acting on the trolley during the impact,
but for the hydrodynamic ones and the reaction of the guide.

\begin{figure}
\centerline{
$a$)
\includegraphics[width=8cm]{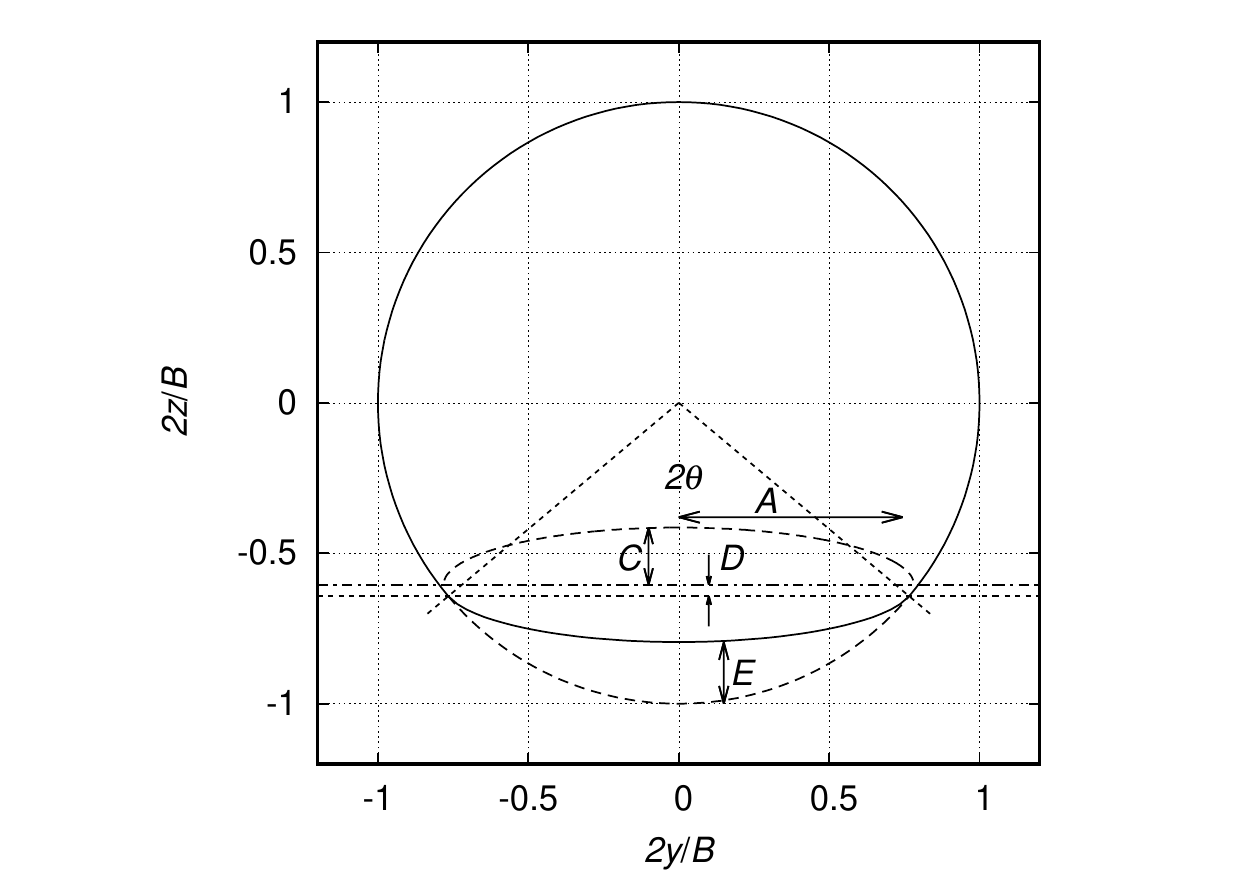}}
\centerline{
$b$)
\includegraphics[width=8cm]{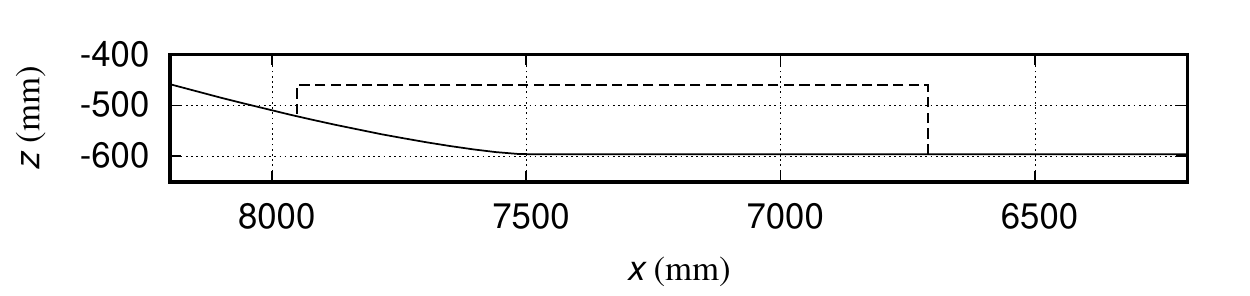}}
\caption{$a$) cylindrical-elliptical cross section; $b$) Longitudinal
section of the specimen: the solid line is the bottom profile of the
fuselage midline whereas the dash lines denote the tested portion
of the fuselage.}
\label{fig:circell}
\end{figure}

The tested specimen are portions of fuselage shapes described by analytical
functions\cite{iafrati2017,iafratiaiaa}.
The fuselage shape considered in this study has a circular-elliptical
cross section (figure \ref{fig:circell}$a$) typical for cargo aircrafts.
The nondimensional cross section contour is defined in terms of two
parameters which are the angle of tangency between the circle and the
ellipse, $\theta$, and the ratio $C/D$. The ellipse semi-axes are
given by
\begin{equation} 
\label{axes}
A=\frac{C}{D} \frac{ \sin \theta }{\sqrt{\left(\frac{C}{D}\right)^2-1} }
\qquad
C=\frac{ A \tan \theta }{\sqrt{\left(\frac{C}{D}\right)^2-1} } \;\;.
\end{equation}
Let
\[
E=1-\cos \theta - C \left(1-\frac{D}{C}\right)
\]
the dimensional equation of the ellipse is
\begin{equation} 
\label{ellipse}
\left(\frac{y/r(x)}{A}\right)^2 + \left(\frac{z/r(x)+1-(E+C)}{C}
\right)^2 = 1 \;\;,
\end{equation}
where $r(x)$ is the radius of the circular portion of the fuselage at the
longitudinal position $x$, origin of the $x$ axis being located at the bow.

Denoting with $B$ the breadth of the main fuselage, the local radius is
given
by
\begin{eqnarray*}
r(x) &=& 0.5 B \sqrt{ 1 - \left(\frac{x-F_B\cdot B}{F_B \cdot B} \right)^2}
\qquad 0 < x < F_B\cdot B \\
r(x) &=& 0.5 B \qquad \qquad \qquad F_B \cdot B < x < x_H \cdot B\\
r(x) &=& 0.5 B + O(x) \qquad \qquad x_H < x < L_B\cdot B
\end{eqnarray*}
where $L_B$ is the total length of the fuselage scaled by the fuselage
breadth, and, similarly, $F_B$ and $R_B$ are the lengths of the forward and
rear portions where the fuselage cross section shrinks with respect to the
main body and $x_H = B(L_B-R_B)$ is the rear end of the main fuselage.
The function $O(x)$ is the offset function which is given by:
\begin{equation} 
\label{sweep}
O(x) = -\frac{B}{E} \sin \left(\frac{x-x_H}{K R_B B}\right) \left[
\frac{x-x_H}{\sin (1/i) R_B B} \right]^{1/i} \;\;,
\end{equation}
where $K$ and $i$ are real coefficients.
The center of the cross section is located at $(x,0,0)$ for $0 < x < x_H$
and at $(x,0,-O(x))$ for $x_H < x < B L_B$.
For the shape considered here the parameters are: $L_B= 7.5$, $F_B=
1.5$, $R_B= 2.5$, $K=1.55$, $i= 2.6$, $C/D= 5$, $\theta = 50$ degrees.
The fuselage breadth for this shape is $B=1500 $ mm and thus, owing to the
limits of the dimensions of the specimen allowed by the trolley,
the fuselage portion used for the tests is that with $ |y| \le 330$ mm
and $ x_I \le x \le x_E$ with $ x_I$ and $ x_E$ being 6710 mm and 7950 mm,
respectively (figure \ref{fig:circell}$b$) .

\begin{figure}
\centerline{\includegraphics[width=8cm]{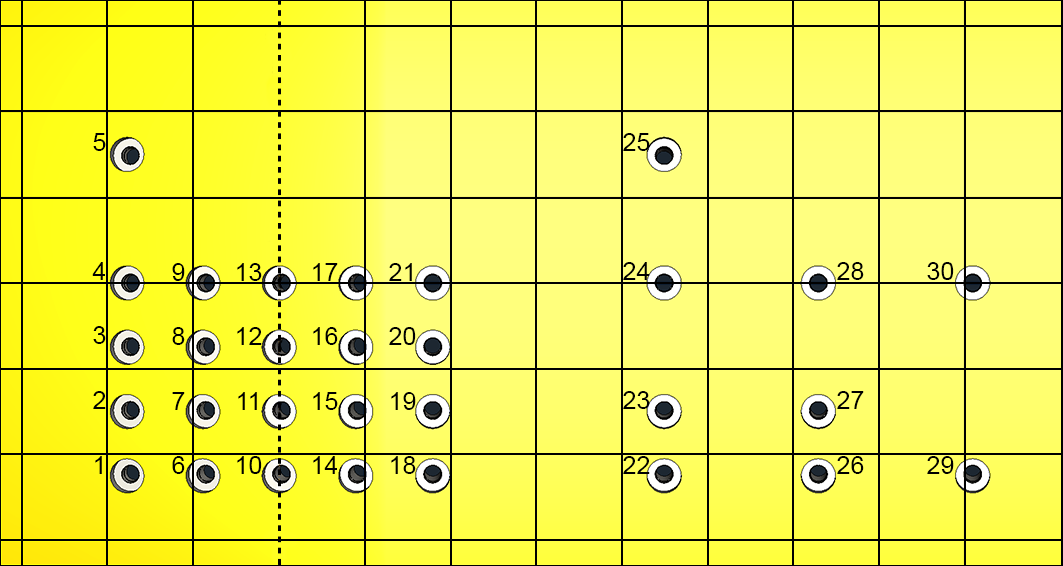}}
\caption{Position of the pressure probes: most of
them are located in the rear part, the trailing edge being on
the left. The dash line passes at the point on the midline
that touches the water surface first at 8 degrees pitch angle whereas the
distance between the solid lines is 100 mm in both directions.}
\label{fig:probes}
\end{figure}

The acquisition box holding the plate is connected to the trolley by a
set of piezoelectric load cells that provide
the loads exerted by the fluid in both longitudinal and normal directions,
the latter measured at rear and front positions.
The rear and front cells are located 70 mm and 1455 mm ahead of the
trailing edge of the specimen, respectively.
A total of 30 pressure probes are installed on the specimen, most of them
located in the rear part behind the 8 degrees contact line (figure
\ref{fig:probes}).
All signals from the sensors are acquired synchronous at a sampling rate
of 200 kS/s, high
enough to capture the pressure peak accurately\citep{vanhuffel2013}.
The on board acquisition is synchronized with two high speed cameras, one
located at the side and used to retrieve the velocity at the impact
and a second one located underwater. The latter, which is operated at 3000
fps, is positioned deep enough to cover all the impact phase, thus
resulting
in a resolution of about 4 mm per pixel.

\section{Results}

The present study is focused on the effects of the horizontal velocity
on the hydrodynamics of the tests performed at a pitch angle of 6 degrees
and vertical to horizontal velocity ratio $V/U=0.0375$.
The analysis of the data is limited to the impact phase, i.e. the phase
lasting from the first contact of the specimen with the water surface up
to the time at which the spray root reaches the leading edge of the
specimen.
After that time, the test speed drops suddenly as a consequence of the
impact with the water on the front face of the specimen and the fore
part of the trolley, and thus the data and the underwater hydrodynamics
become too much dependent on the shape of the upper side of the acquisition
box.
As done in Ref. \onlinecite{iafrati2016}, the end of the impact phase 
is estimated from
the propagation velocity computed on the basis of the times of the pressure
peak passages at probe P28 and P30.

\begin{figure*}
\centerline{
$a)$
\includegraphics[width=80mm]{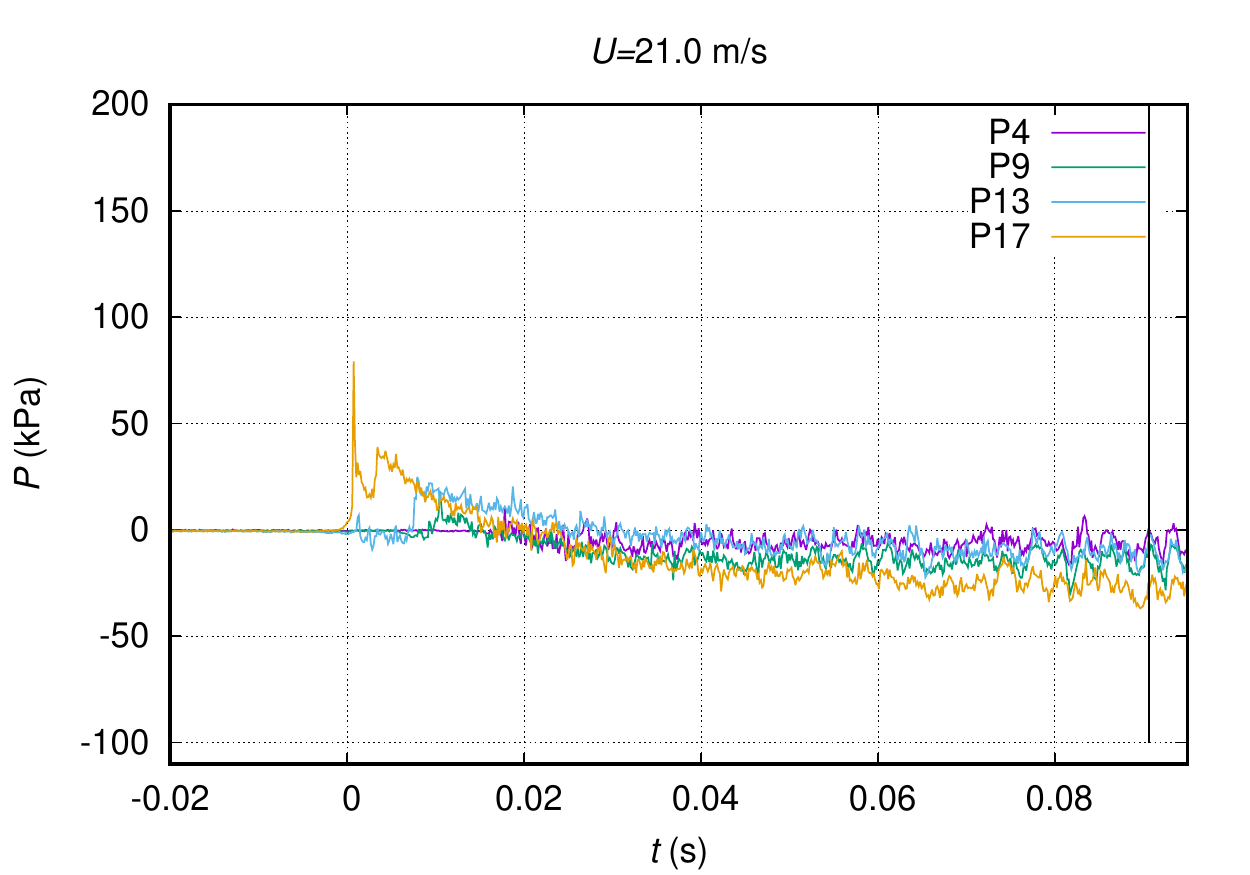}
$b)$
\includegraphics[width=80mm]{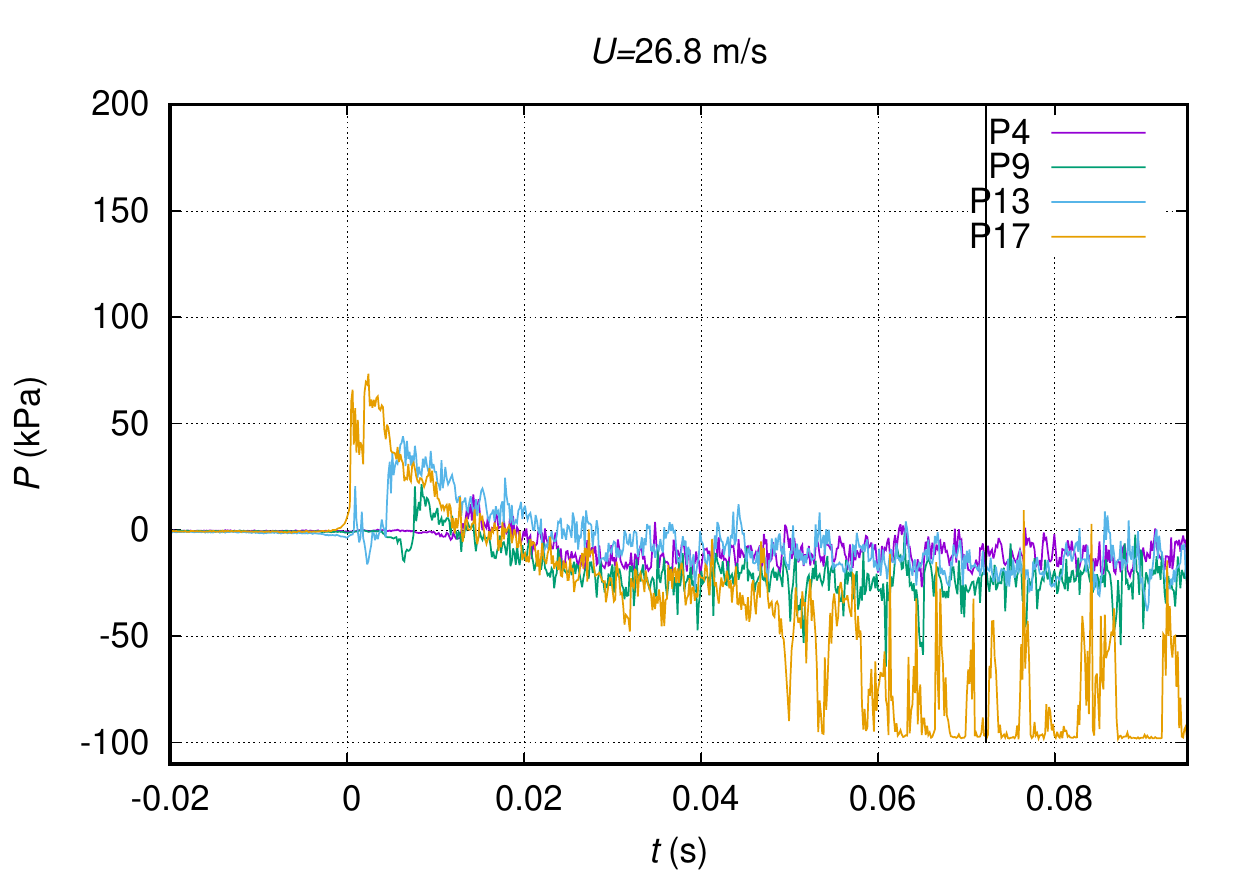}
}
\centerline{
$c)$
\includegraphics[width=80mm]{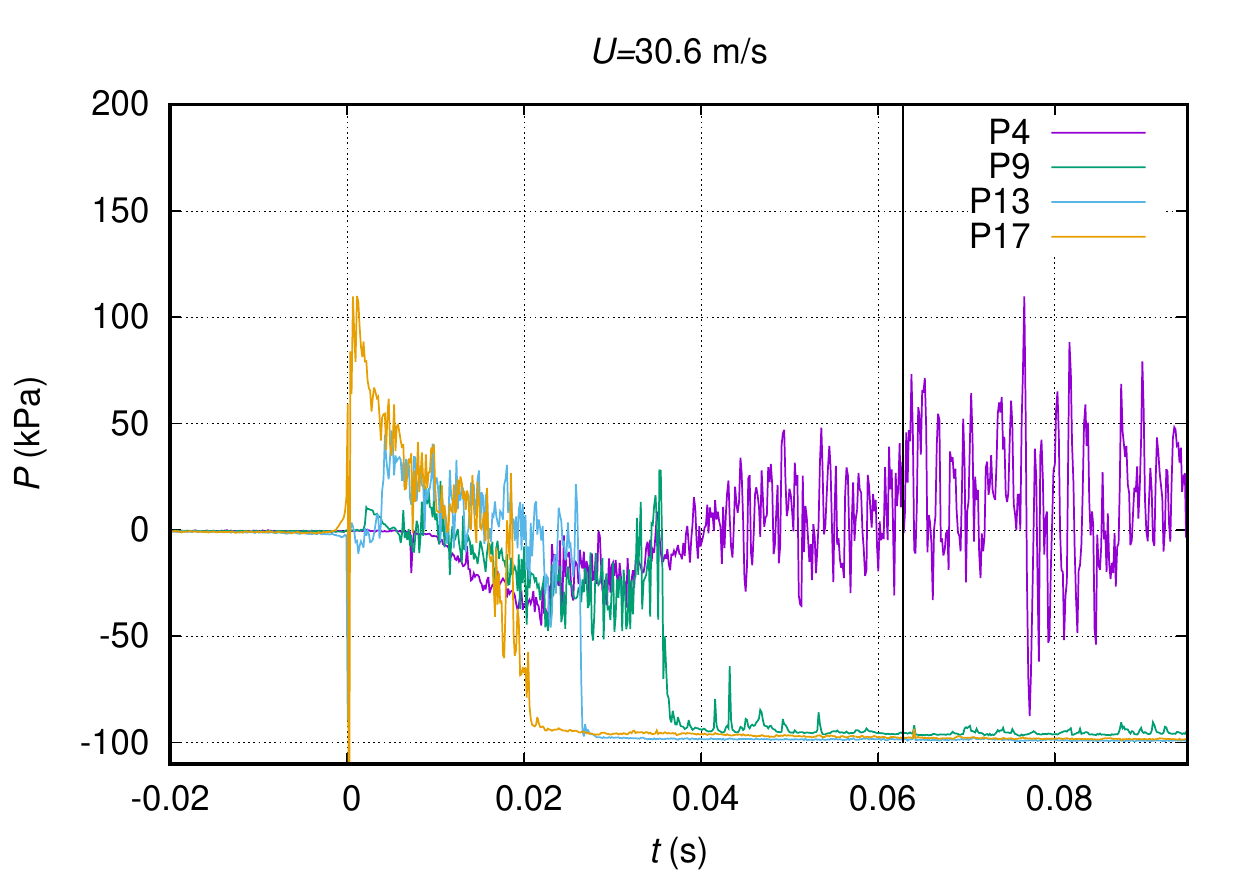}
$d)$
\includegraphics[width=80mm]{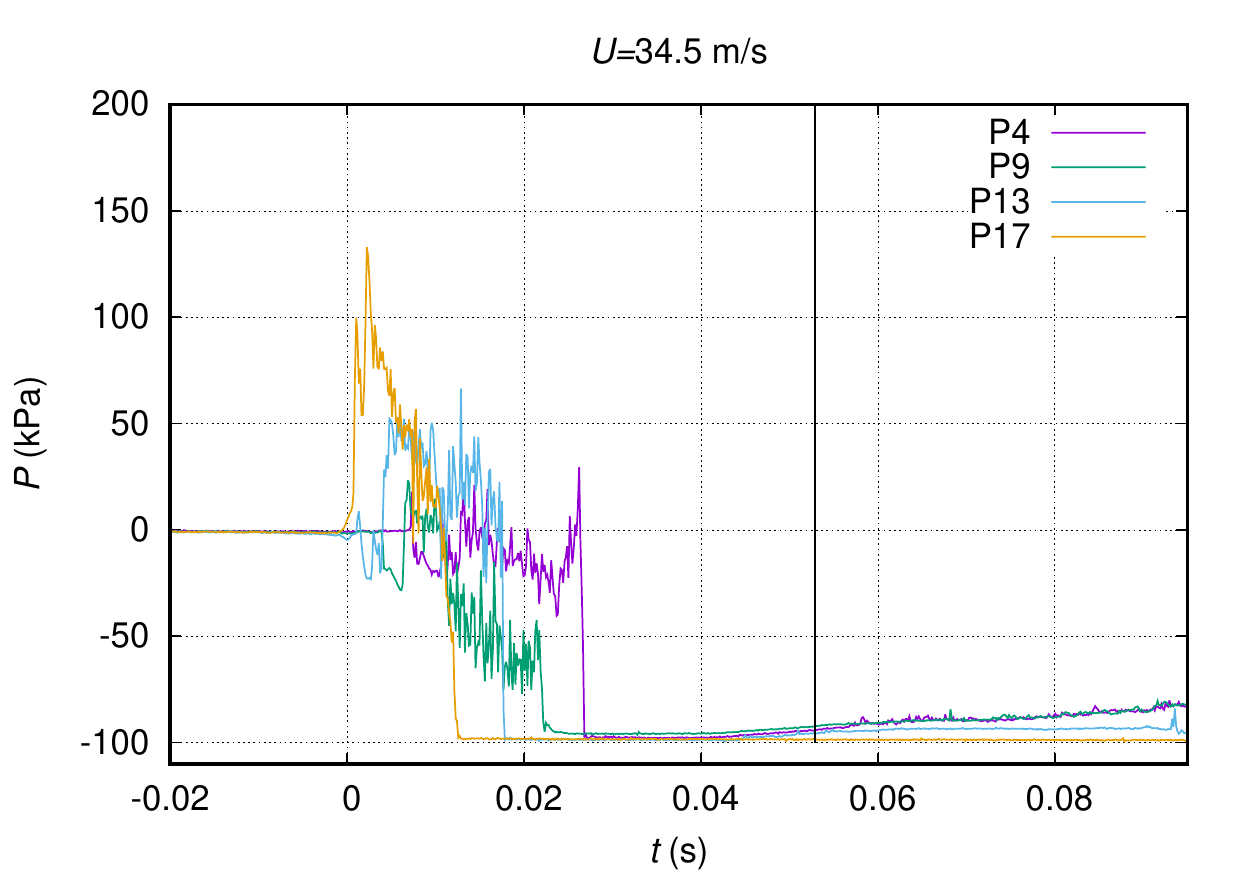}
}
\centerline{
$e)$
\includegraphics[width=80mm]{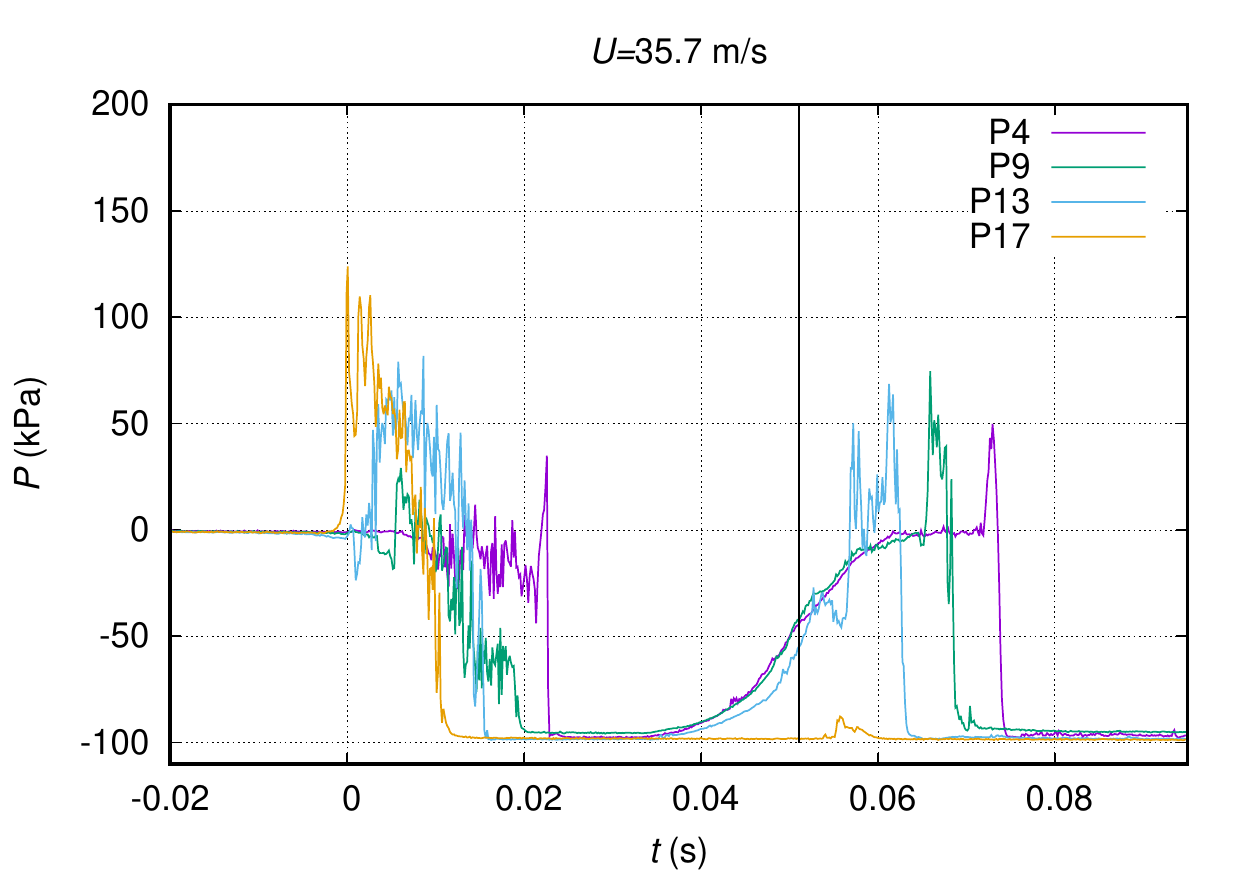}
$f)$
\includegraphics[width=80mm]{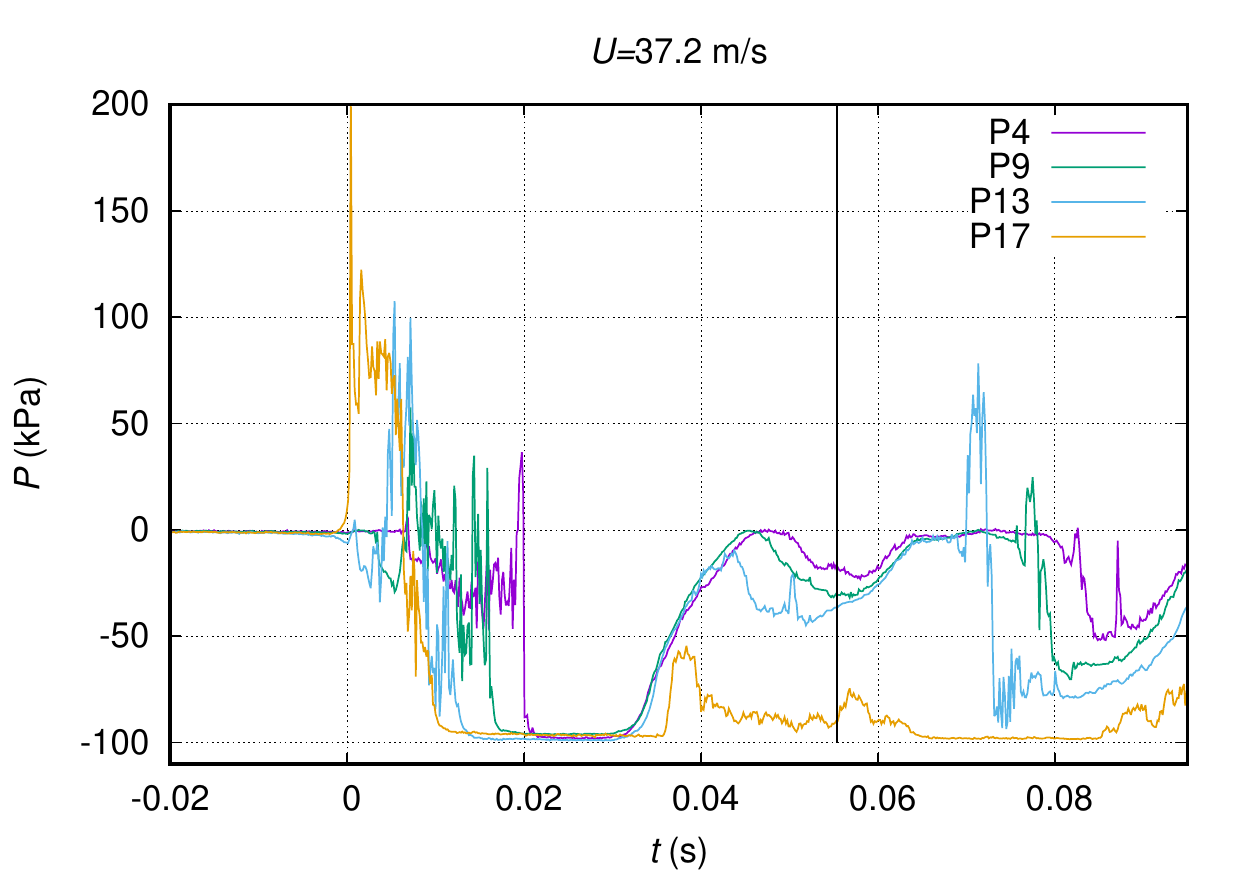}
}
\centerline{
$g)$
\includegraphics[width=80mm]{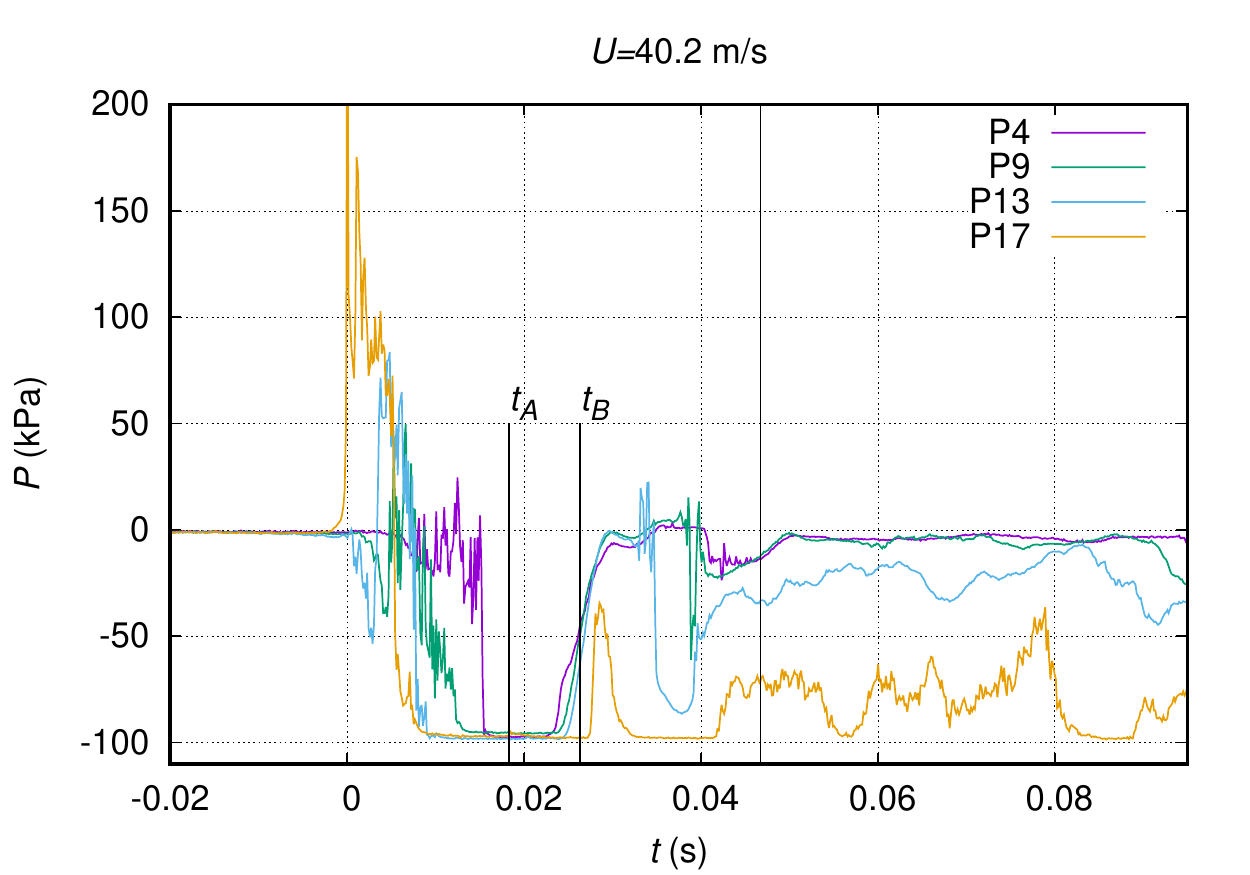}
$h)$
\includegraphics[width=80mm]{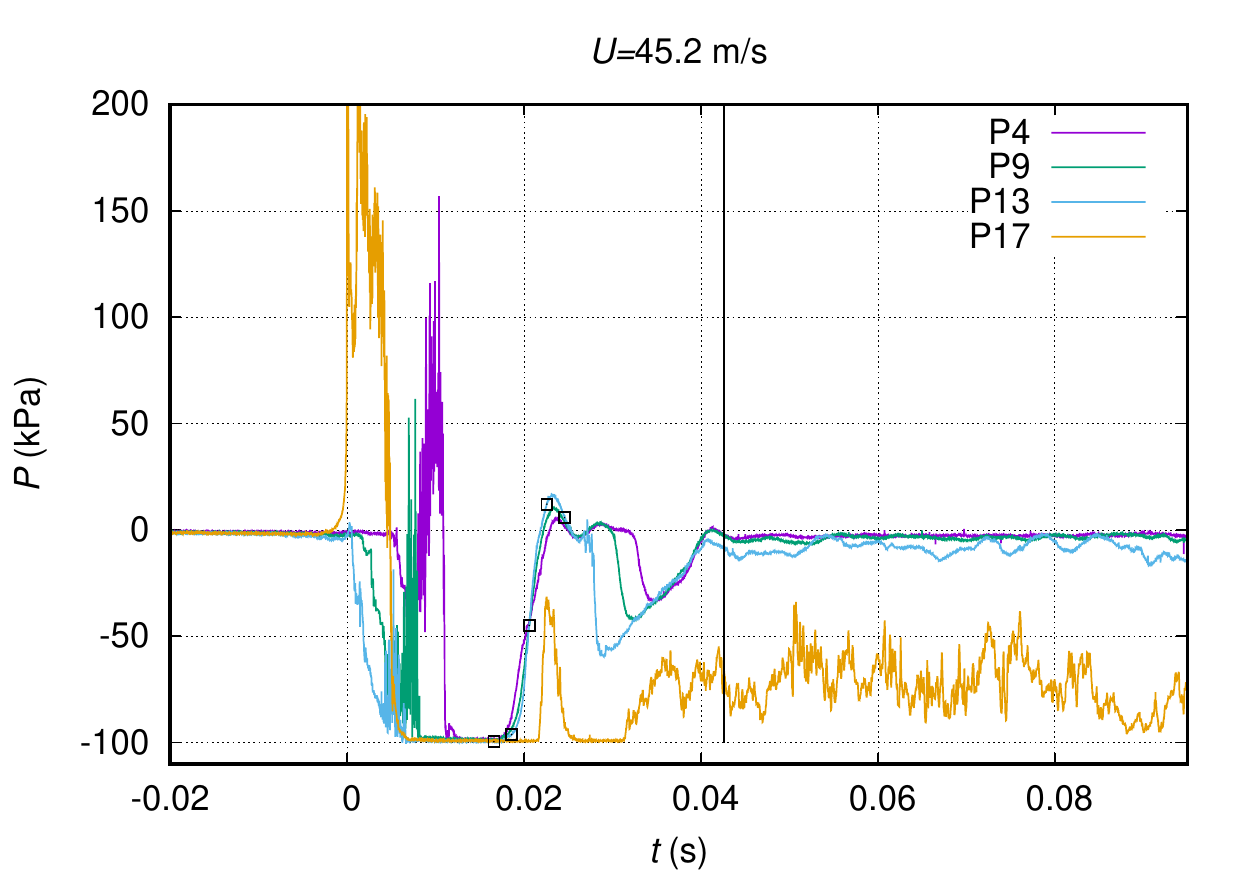}
}
\caption{Time histories of the relative pressure measured at probes located
along
the midline in the rear part of the specimen. The vertical lines are
located at the time when the spray root reaches the leading edge of the
specimens, which is the end of the impact phase.
On $g)$ the empty boxes drawn on P13 corresponds
to the times of the underwater images shown in figure
\ref{fig:cav_ventil}.}
\label{fig:press_vel}
\end{figure*}

The analysis is started from the signals of the four pressure probes
located at the rear, along the specimen midline.
By combining the pressure data with the underwater images, different
cavitation/ventilation modalities are observed when varying the impact
velocity, with the transition from one modality to the other occurring
within
velocity variations of few m/s (Table \ref{tab:cavvent}).
The time histories of the pressures recorded at different impact velocities
by the four probes located along the midline in the rear portion of the
specimen are shown in figure \ref{fig:press_vel}.
It is worth noticing that all the data presented in the following are
intentionally unfiltered.
The data provide the relative pressure, the ambient pressure being used as
reference value.
The water temperature during the tests was between 18$^\circ$C and 
21$^\circ$C. At such
temperatures the absolute vapor pressure is in the range 2.0 to 2.7 kPa,
which correspond to relative values of -99.30 kPa and -98.60 kPa,
respectively.

In all cases, at the initial contact, probe P17 exhibits a sharp rise up
and
a pressure peak.
A sharp rise is also found on P13 and P9, although the peak intensities are
lower and are delayed with respect to P17 due to the time needed for the
spray root to propagate.
Next, the probes display a gentle pressure reduction and negative pressures
are observed as well.

The occurrence of a pressure peak at P17 and P13 followed by the negative
pressure can be explained with the asymmetry of the body shape and of
the relative velocity.
As shown in Ref. \onlinecite{semenov2006}, in the case of asymmetric impact the
pressure peak on the side with reduced deadrise angle is much higher than
that on the other side.
At the beginning of the entry process, the two peaks are close to each
other and the pressure level in between is also high. However,
as the body penetrates the two peaks move away and the pressure at the apex
drops substantially. The pressure at the apex diminishes even more due to
the large difference in the intensity of the pressure peaks which
leads to a cross flow about the apex that generate negative pressures
\cite{riccardi2004} and, depending on the conditions, it may eventually
lead to cavitation or ventilation \citep{judge2004}.
Another interpretation of the phenomenon based on the flat plate solution
is provided in the next section.

At $u = 21.0$ m/s (figure \ref{fig:press_vel}$a$), it is seen that indeed
the pressure at all four probes turns negative, with P17, being very
close to the sharp change in the shape, reaching the lowest values.
However, the pressure measured by all probes is always much above the vapor
pressure.

When increasing the test speed to 26.8 m/s, there is a corresponding growth
of the pressure peaks at P9, P13 and P17 as well as a reduction in
the negative values attained in the later stage.
Shortly before the end of the impact phase, the pressure at P17 exhibits
sudden drops to the vapor pressure indicating that cavitation
is about to start (figure \ref{fig:press_vel}$b$).
By looking at the corresponding underwater images (figure
\ref{fig:undw_img}$c,d$) there is no clear evidence of cavitation and very
likely cavitation is only local and it is associated with the interaction
of
the flow with the pressure probes and with their installation on the
specimen,
as explained in Ref. \onlinecite{iafrati2015}.

The occurrence of cavitation becomes evident at $U=30.6$ m/s, when P17, P13
and P9 exhibit sharp pressure drops remaining for long time intervals at
the
vapor pressure value (figure \ref{fig:press_vel}$c$).
The vapor cavity is indeed visible from the underwater images already at
the middle of the impact phase (figure \ref{fig:undw_img}$e$).
However, differently from what happens at higher speeds, the cavity does
not propagate much beyond P9 (figure \ref{fig:undw_img}$f$).
For such reason this is considered as an Inception Cavitation case.

It is worth noticing that whereas before the sharp drops the pressure
signals are characterized by large and high-frequency fluctuations, which
disappear once the vapor pressure is reached.
The pressure fluctuations are a consequence of the irregularities in the
relative motion between the body and the water surface caused by the
structural vibrations of the trolley and of the guide together with the
added mass of the water.
Once the cavitation starts, the pressure inside the cavitation bubble is
constant and the irregularities in the vertical motion are absorbed by the
deformations of the cavity, whereas the added mass of the vapor is
negligible.

More quantitatively, the phenomenon is highlighted by the continuous
wavelet transform of the pressure signals, as that provided in figure
\ref{fig:wavelet}. Whereas the signals are rather broad banded in the early
stage and in the pressure drop, there are no frequencies appearing
in the cavitation phase.

\begin{table}
\begin{center}
\begin{tabular}{@{}|c|c|c|c|c|c|c|c|c|} \hline \hline
$U$ (m/s) & $21.0$ & $26.8$ & $30.6$ & $34.5$ & $35.7$ & $37.2$ & $40.2$
& $45.2$\\
Type &  NC & NC & IC  & FCB  & CAV & CV & CV & CV \\ \hline \hline
\end{tabular}
\end{center}
\caption{Cavitation/ventilation conditions found when varying the
horizontal speed: NC: No Cavitation; IC: Incipient Cavitation; FCB: Fully
Cavitating Bubble; CAV: Cavitation Alternate with Ventilation;
CV: Cavitation followed by Ventilation.}
\label{tab:cavvent}
\end{table}

When increasing the horizontal velocity at the impact to 34.5 m/s, the
cavity propagates till the backward step of the specimen, as it is seen
from both the pressure measurements (figure \ref{fig:press_vel}$d$) and the
underwater images in figures \ref{fig:undw_img}$g,h$. Figure
\ref{fig:undw_img}$h$ also displays the initial formation of the ventilated
cavity but it remains confined to a limited portion at the rear of the
specimen.

A further increase of the horizontal velocity of about one meter per second
is already enough for the ventilation effects to propagate forward up to
P13 at
least. The pressure in the bubble rises quite gently in this case (figure
\ref{fig:press_vel}$e$).
In such
specific test the ambient pressure is reached shortly after the end of the
impact phase. All probes, but for P17, display a positive pressure
peak and next there is the development of a new cavitating bubble.
The underwater picture \ref{fig:undw_img}$j$, being taken at the end of the
impact phase, shows the ventilated cavity only and not the formation of the
new cavitation bubble.

The transition to ventilation becomes stable when increasing the
horizontal velocity above 37.2 m/s. As shown in figures
\ref{fig:press_vel}$f,g,h$, the ventilated cavity propagates till P17.
However, sensor P17 is just behind the lowest point of the geometry and
in the sharply curved region so that, also due to the intense cross flow,
never gets to the fully ventilated conditions.

By comparing the time histories of the pressures for the different cases it
is seen that the time duration of the cavitating phase shrinks with
increasing the horizontal velocity and the development of the ventilation
is anticipated to less than half of impact phase.
Whereas the delay of the pressure drops of the different probes indicates
that the cavitating bubble needs some time to reach the different probes,
when the bubble reaches the probe the pressure drops suddenly. Differently
from that, when the ventilation occurs the pressure rises at a given level
at all probes at the same time, and it is instead the growth rate of the
pressure that increases with increasing the test speed.
This is because when the cavitating bubble gets in contact with the air at
the rear of the specimen, the higher pressure is felt everywhere within the
bubble. However, the pressure in the bubble does not
jump instantaneously to the ambient value but it grows
as the ventilation effects propagate forward into the bubble.
The propagation of ventilation effects is highlighted by the
sequence of underwater images provided in figure \ref{fig:cav_ventil} which
correspond to the boxes in the time histories of the pressure shown in
figure \ref{fig:press_vel}$h$.

The propagation of the cavitation and ventilation bubbles can be further
inferred from figure \ref{fig:press_contours} where the pressure contours
at two different times during the impact phase of the test at 40.2 m/s are
drawn over the corresponding underwater images.
As the probes are located only on one side of the specimen, the symmetry of
the pressure is exploited.
In spite of the limited number of probes, the pressure contours are well
overlapped to the cavitation bubble. Somewhat more noisy are the contours
in
the ventilation phase, but it is still possible to recognize the
propagation
of the front.

Whereas the analysis of the pressures is essential to understand the
hydrodynamic phenomena taking place beneath the specimen, their practical
consequences
are more understandable from the time histories of the normal forces, and
particularly by the comparison of the forces measured at the rear and at
the front, provided in figure~\ref{fig:nor_for_vel}.
The data clearly show that until ventilation appears around 35 m/s, the
force at the rear is more and more negative as the horizontal
velocity is increased, although the total force keeps increasing of course.
However, as soon as the ventilation enters into play a sudden increase in
the
force at the rear is observed, which is concurrent with the pressure
growth. This is made evident at the highest test speed where it can be
noticed that the marks placed along the pressure rise (figure
\ref{fig:press_vel}$g$) perfectly correlate with the sharp
growth of the force measured at the rear (figure \ref{fig:nor_for_vel}$g$).

\begin{figure*}
\centerline{
$a)$
\includegraphics[width=52mm,angle=-90]{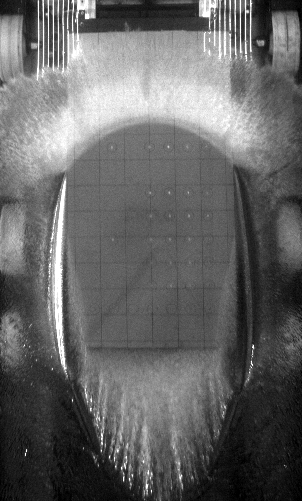}
$b)$
\includegraphics[width=52mm,angle=-90]{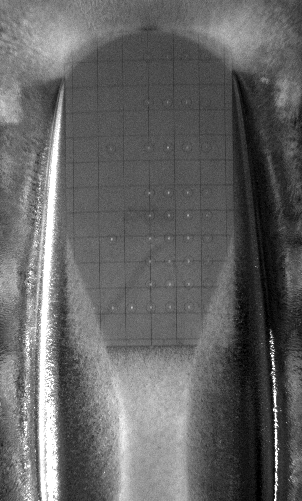}
}
\centerline{
$c)$ \includegraphics[width=52mm,angle=-90]{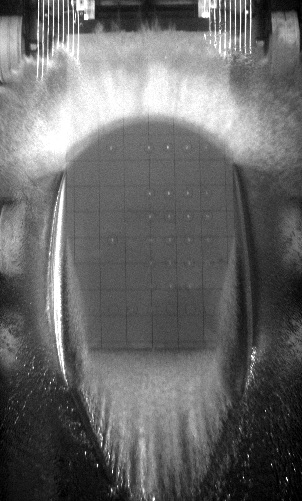}
$d)$ \includegraphics[width=52mm,angle=-90]{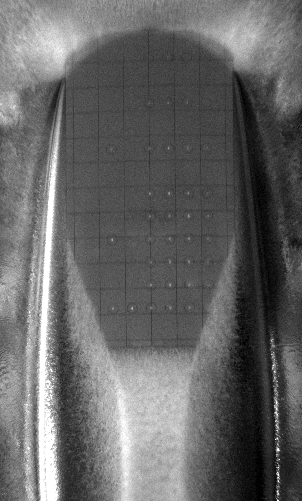}
}
\centerline{
$e)$ \includegraphics[width=52mm,angle=-90]{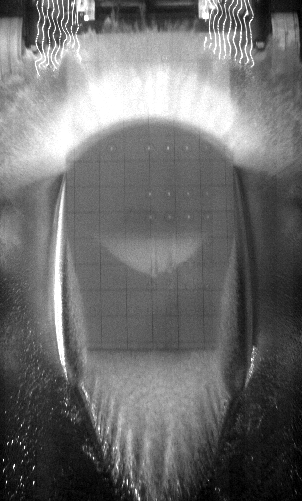}
$f)$ \includegraphics[width=52mm,angle=-90]{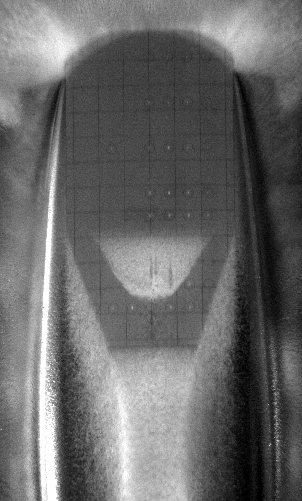}
}
\centerline{
$g)$ \includegraphics[width=52mm,angle=-90]{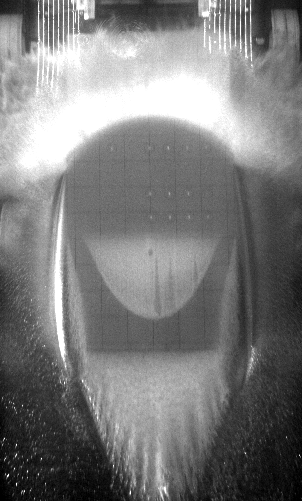}
$h)$ \includegraphics[width=52mm,angle=-90]{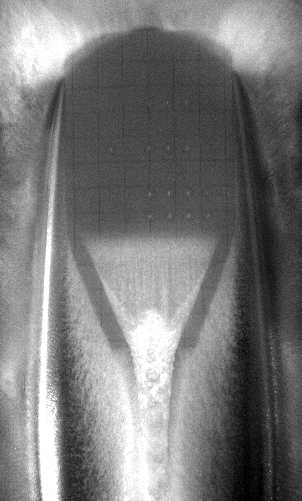}
}
\caption{See caption on the next page}
\end{figure*}

\setcounter{figure}{3}
\begin{figure*}
\centerline{
$i)$ \includegraphics[width=52mm,angle=-90]{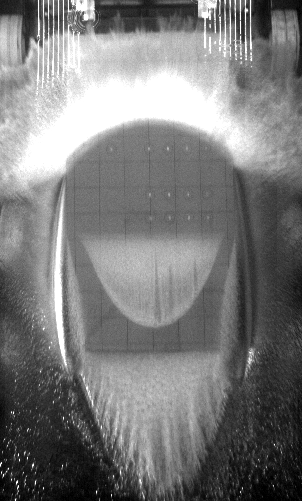}
$j)$ \includegraphics[width=52mm,angle=-90]{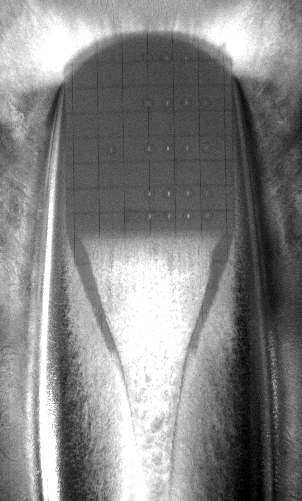}
}
\centerline{
$k)$ \includegraphics[width=52mm,angle=-90]{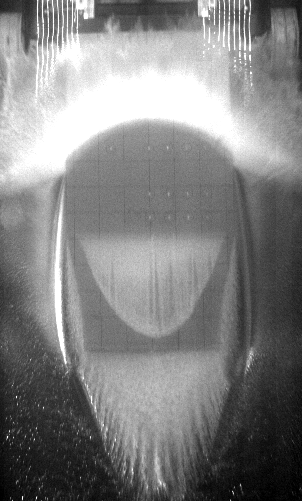}
$l)$ \includegraphics[width=52mm,angle=-90]{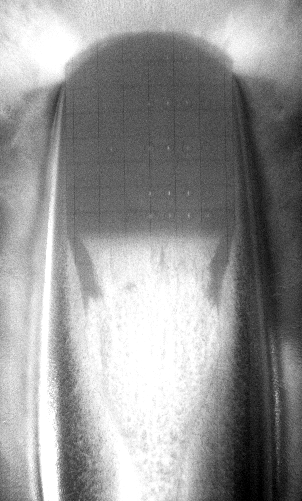}
}
\centerline{
$m)$ \includegraphics[width=52mm,angle=-90]{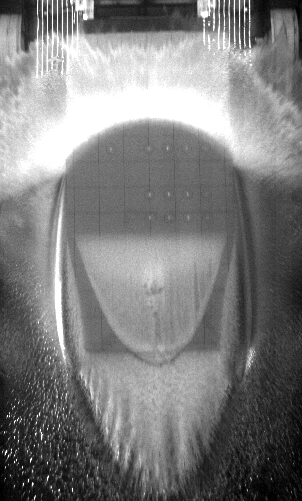}
$n)$ \includegraphics[width=52mm,angle=-90]{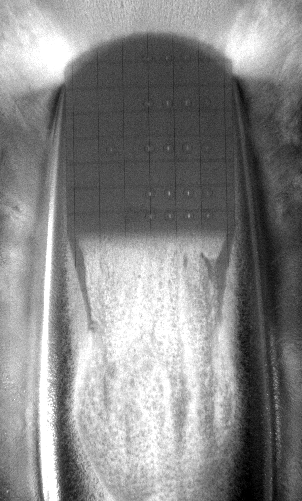}
}
\centerline{
$o)$ \includegraphics[width=52mm,angle=-90]{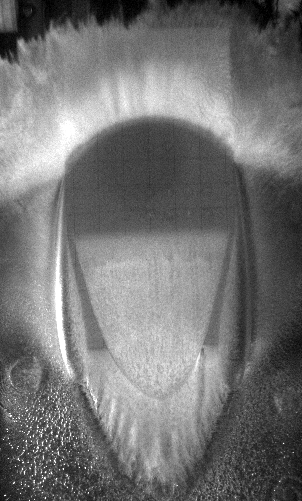}
$p)$ \includegraphics[width=52mm,angle=-90]{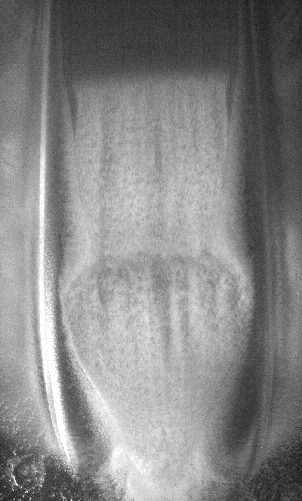}
}
\caption{Underwater images of the tests. For each condition, the pictures
at half and at the end of the impact phase are provided on the left and
right columns, respectively.
Test conditions refer to: $a,b)$ 20.9 m/s; $c,d)$ 26.8 m/s; $e,f)$
30.6 m/s; $g,h)$ 34.5 m/s; $i,j)$ 35.6 m/s; $k,l)$ 37.2 m/s; $m,n)$
40.2 m/s; $o,p)$ 45.2 m/s. }
\label{fig:undw_img}
\end{figure*}

\begin{figure}
\centerline{
\includegraphics[width=8cm]{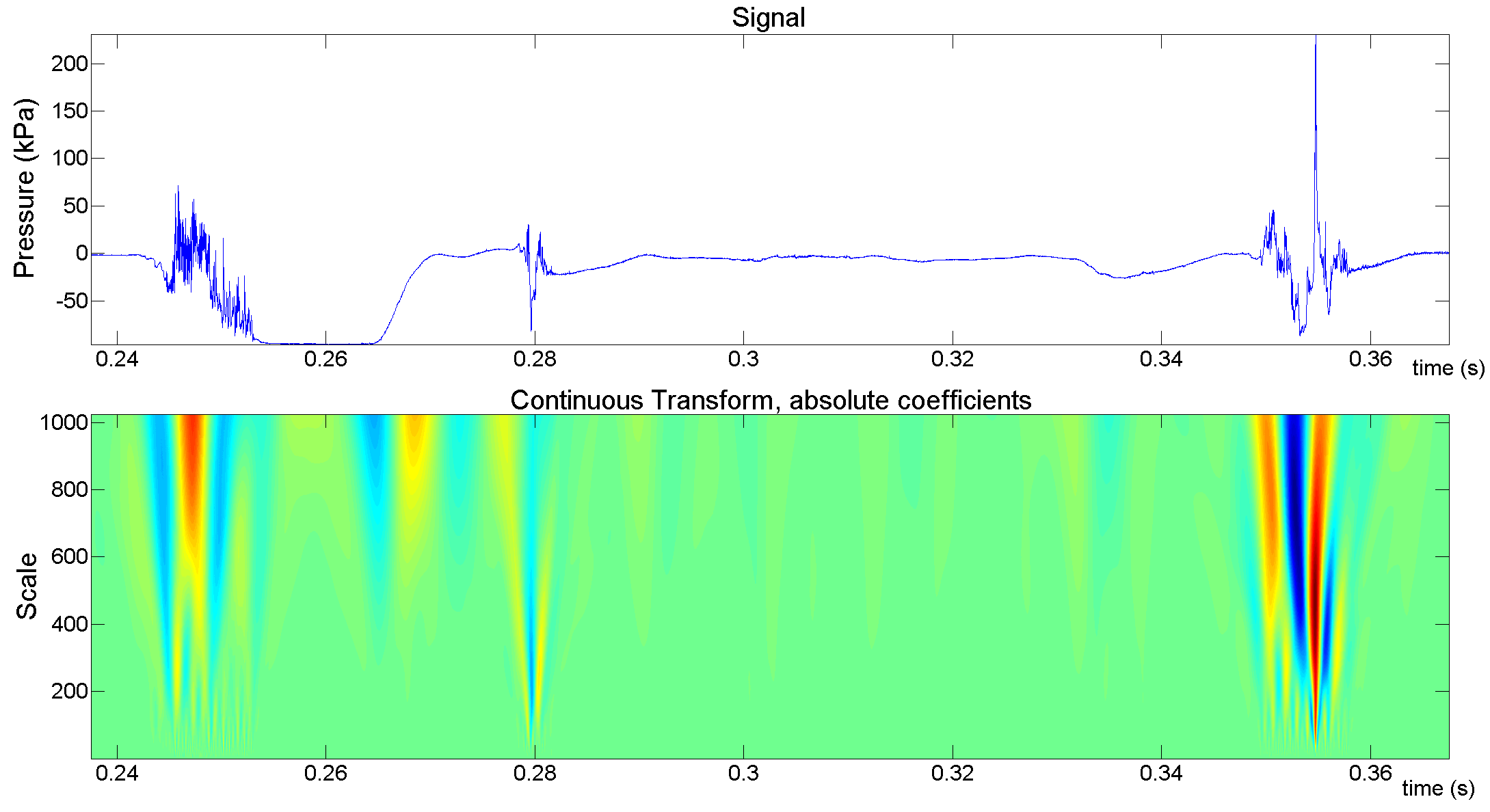}}
\caption{Time history of the pressure recorded at probe P9 in the test at
40.2 m/s ({\em top}) and of the corresponding continuous wavelet transform
({\em bottom}). The results clearly highlight the suppression of the
pressure fluctuations during both the cavitation and ventilation phases.
}
\label{fig:wavelet}
\end{figure}

Results in figure \ref{fig:nor_for_vel} indicate that the ventilation
also leads to an increase of the total loading, which is to say that
because of the ventilation, the change in the pressure distribution at the
rear of the specimen does not only affect the position of
the center of loads but also the total force on the specimen.
The sudden changes in the pressure distribution and their consequences
in terms of the center of loads discussed above are remarkably different
from what found for the flat plate in Ref. \onlinecite{iafrati2016}, 
where a much
smoother behavior of the loads and a monotonic  motion of the center
of loads, arriving up to about 80 $\%$ of the plate length, was found.

The sudden changes in the hydrodynamics and the subsequent effects on the
loading distribution which are a consequence of the fuselage curvature and
of the high horizontal speed, are expected to have important effects on
the aircraft dynamics and have to be correctly reproduced in the ditching
analyses.

\begin{figure}
\centerline{
$a$) \includegraphics[width=8cm]{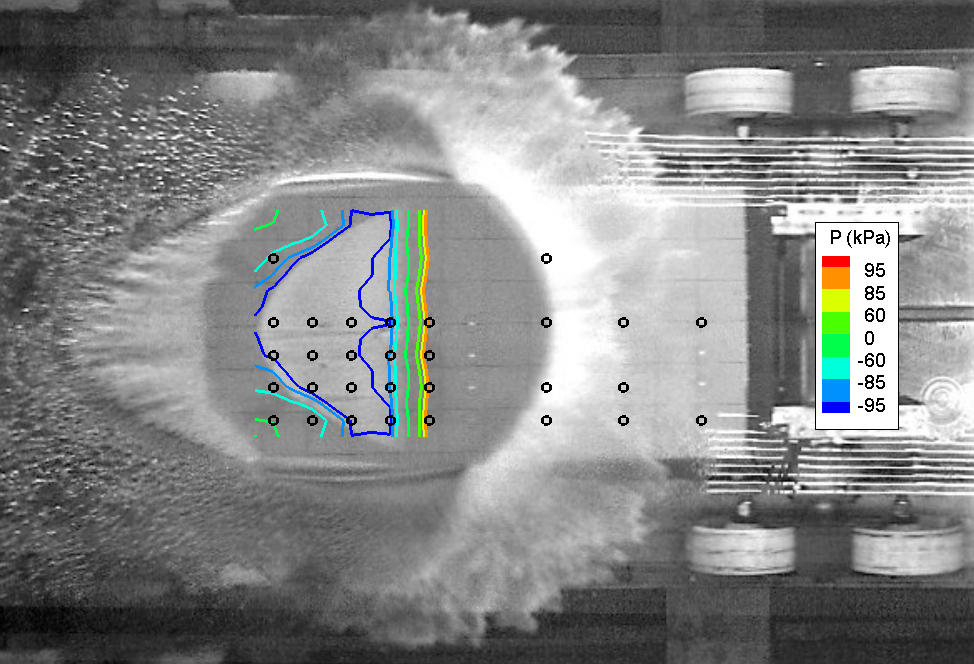}}
\centerline{
$b$) \includegraphics[width=8cm]{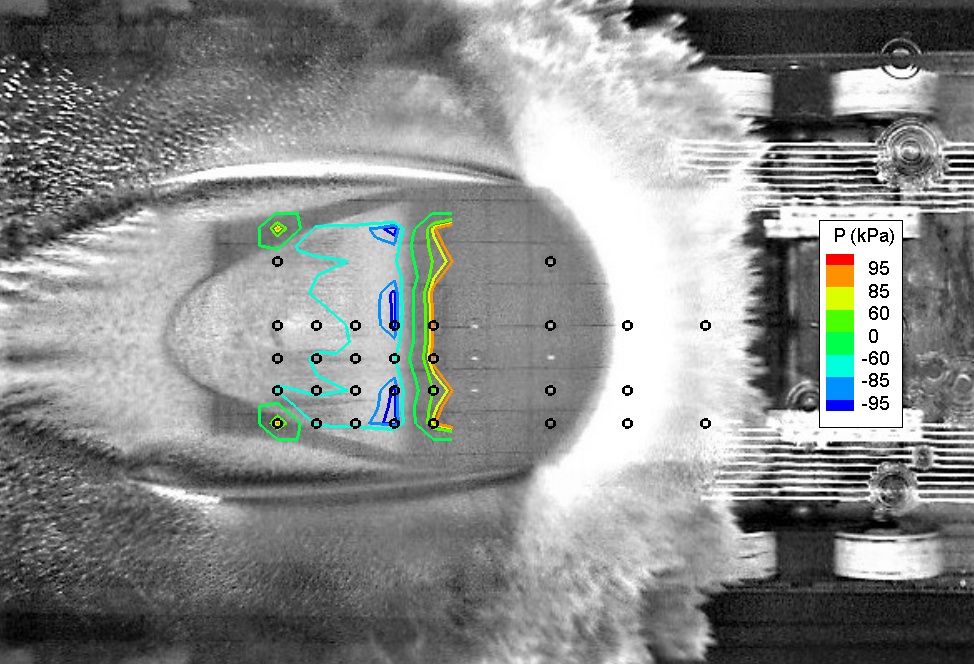}}
\caption{Pressure contours drawn at $t= t_A$ ({\em top}) and $t=t_B$
({\em bottom} for the test at 40.2 m/s (see figure \ref{fig:press_vel}$g$).
}
\label{fig:press_contours}
\end{figure}

\section{Discussion}

The above results show what is actually happening and how the phenomena
change when increasing the horizontal velocity but they do not explain
the reasons.
Without additional considerations, the message could be misleading as it is
not the horizontal velocity itself that triggers the phenomena but it is
its combination with the body shape and with the longitudinal
curvature in particular.
Unfortunately, the problem cannot be easily represented by a simple model
and a full three-dimensional flow solver capable of describing the
nonlinear free surface dynamics would be necessary.
Such high fidelity models are not available at this time, which
is one of the motivations for the present study, and thus some
attempts to explain the phenomenon can only be done based on approximate
representations.

\begin{figure*}
\centerline{
$a$) \includegraphics[width=80mm]{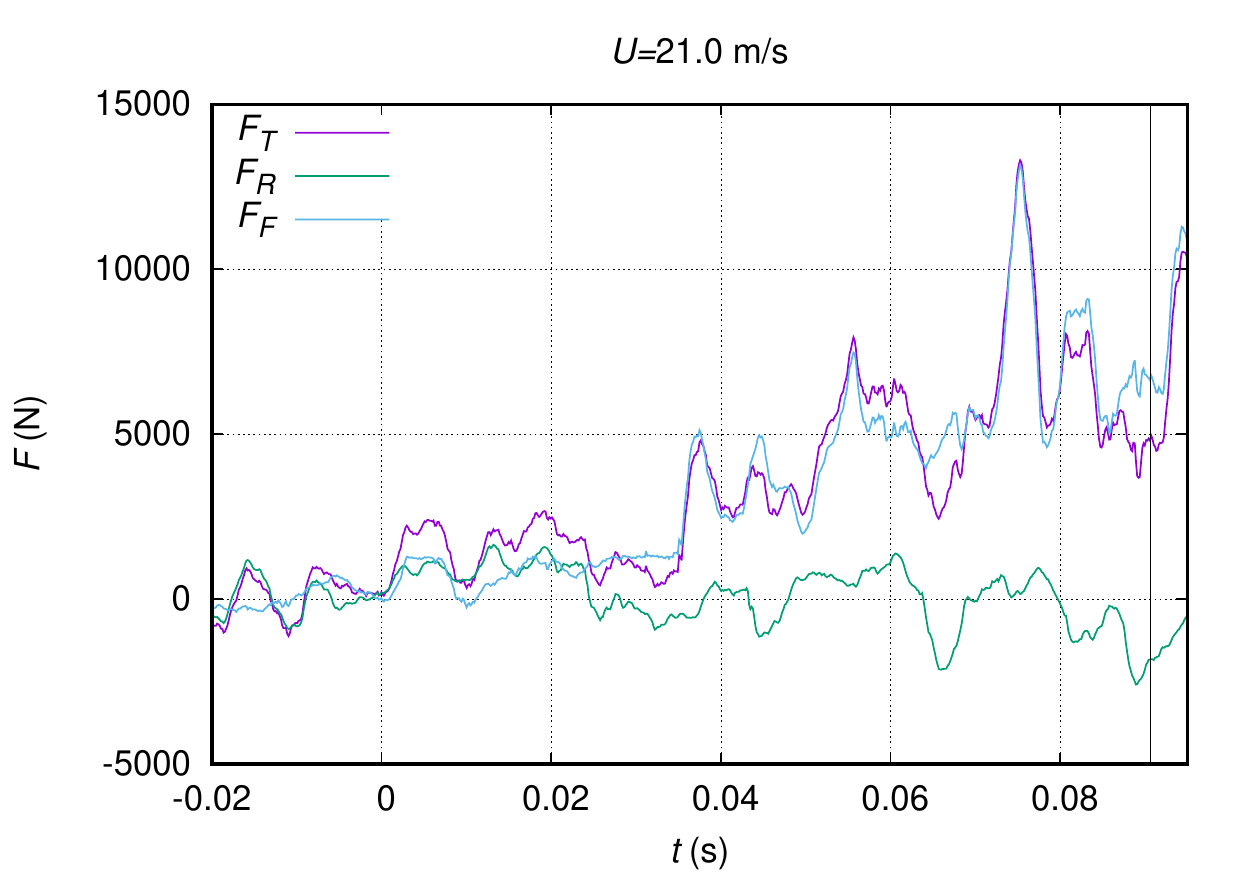}
$b$) \includegraphics[width=80mm]{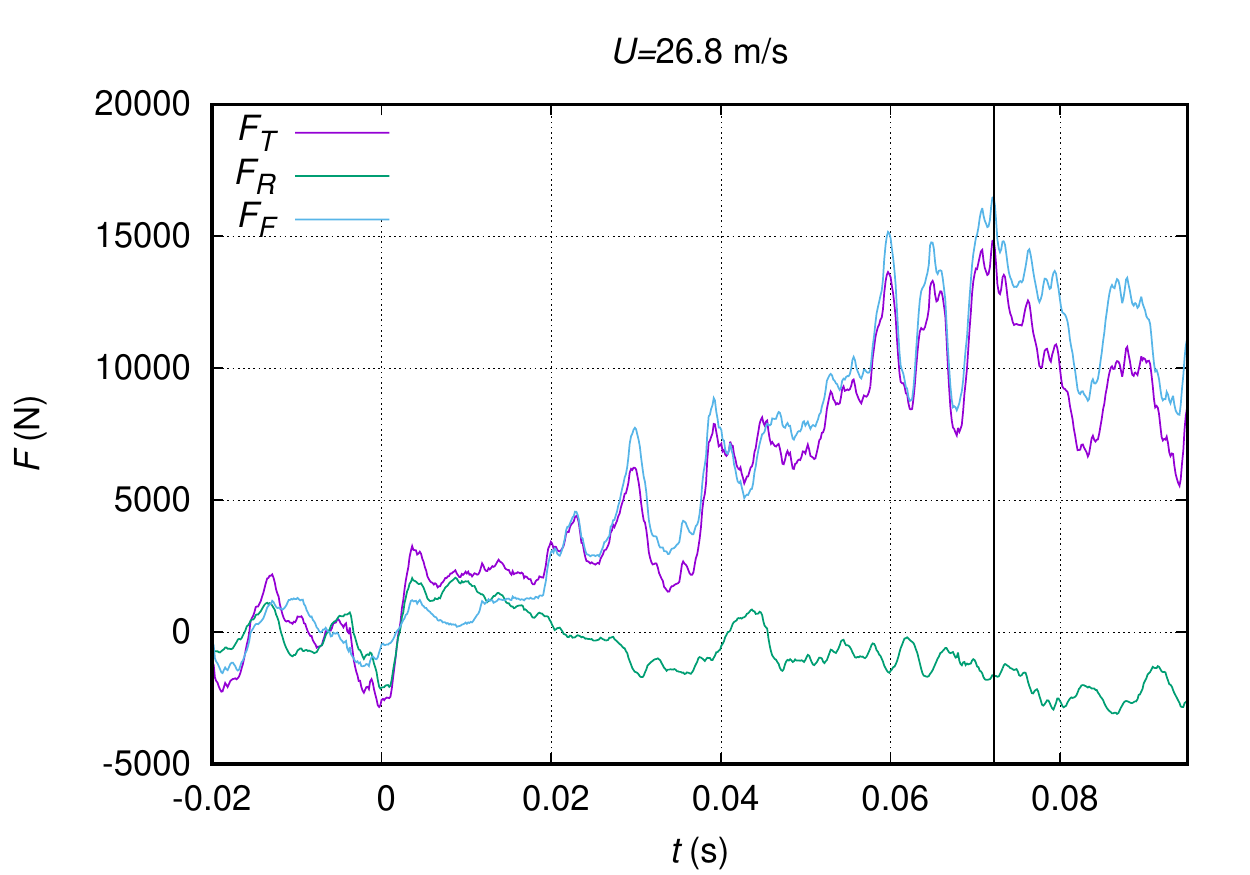}
}
\centerline{
$c$) \includegraphics[width=80mm]{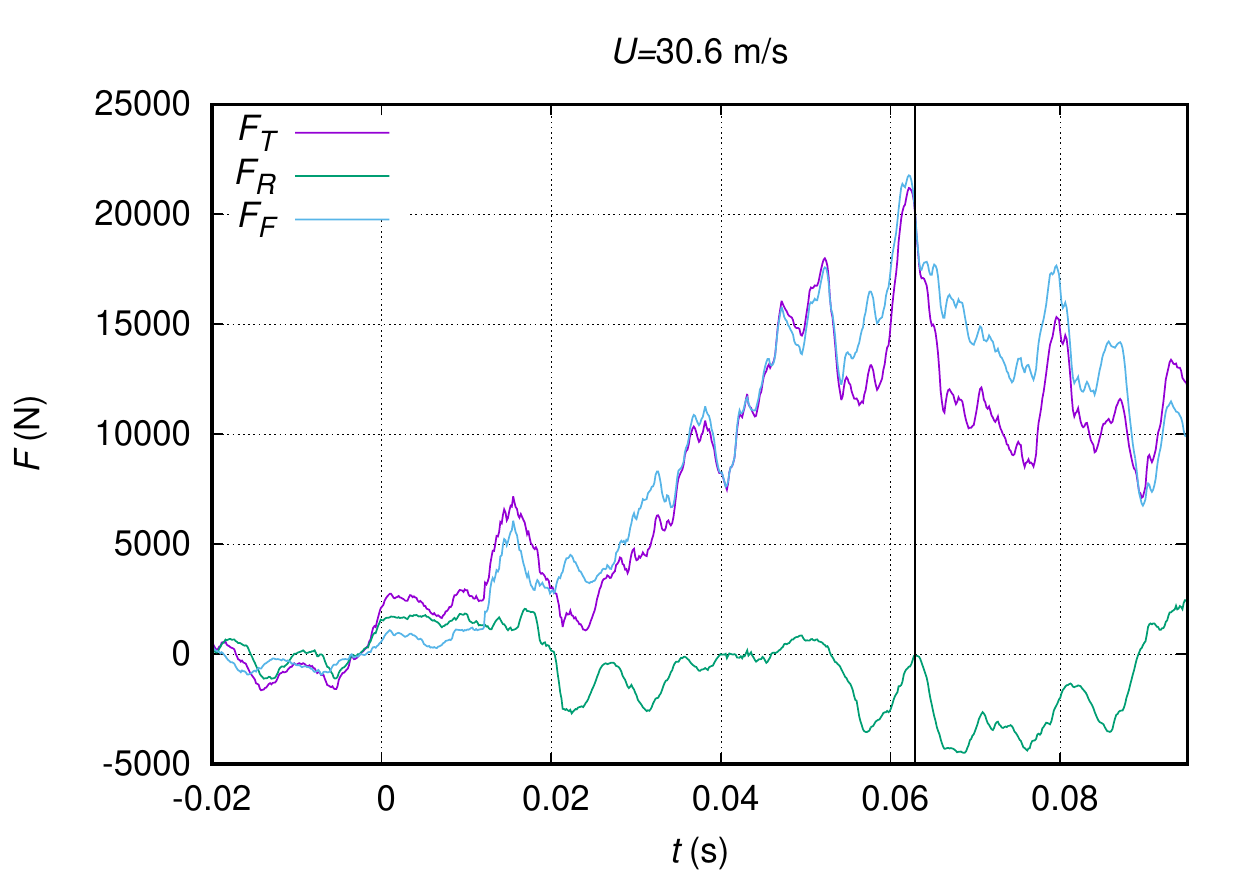}
$d$) \includegraphics[width=80mm]{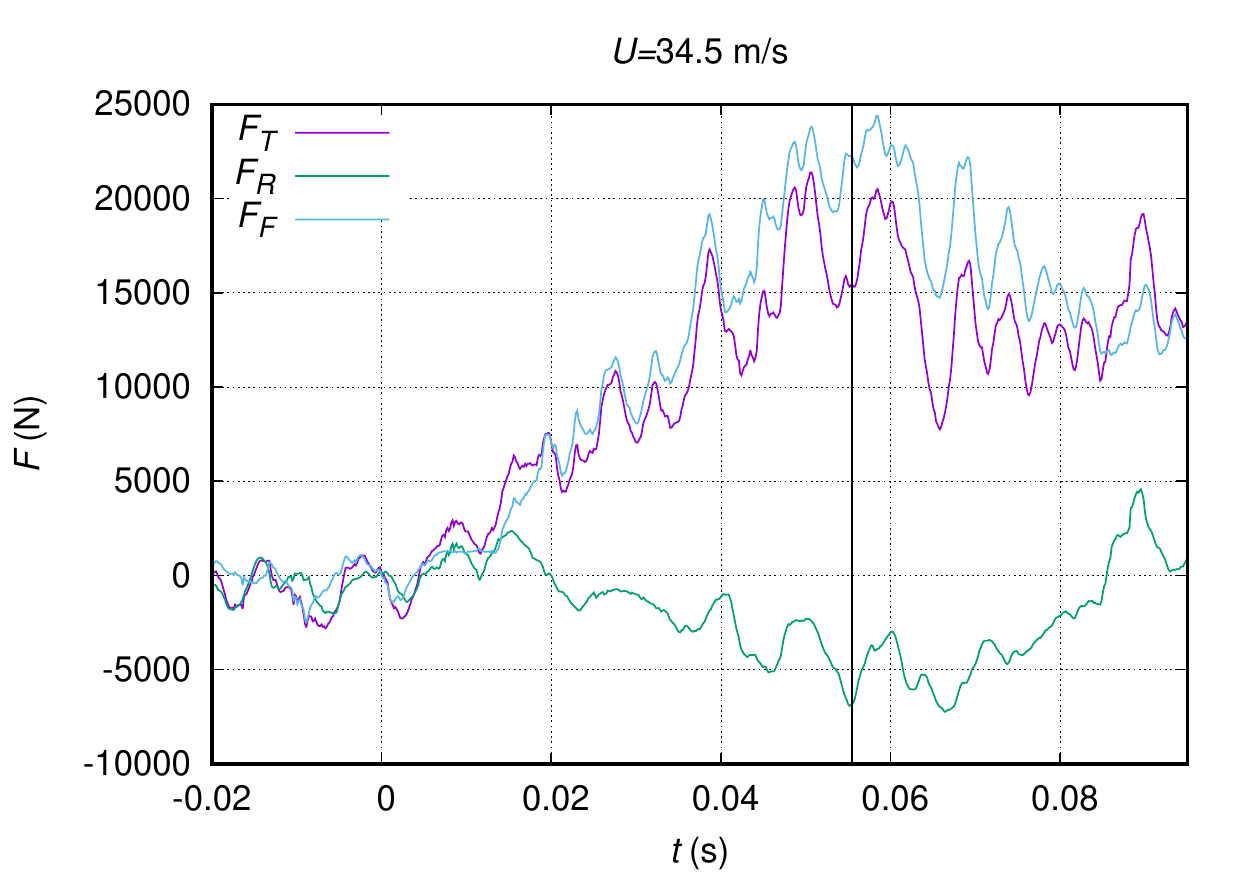}
}
\centerline{
$e$) \includegraphics[width=80mm]{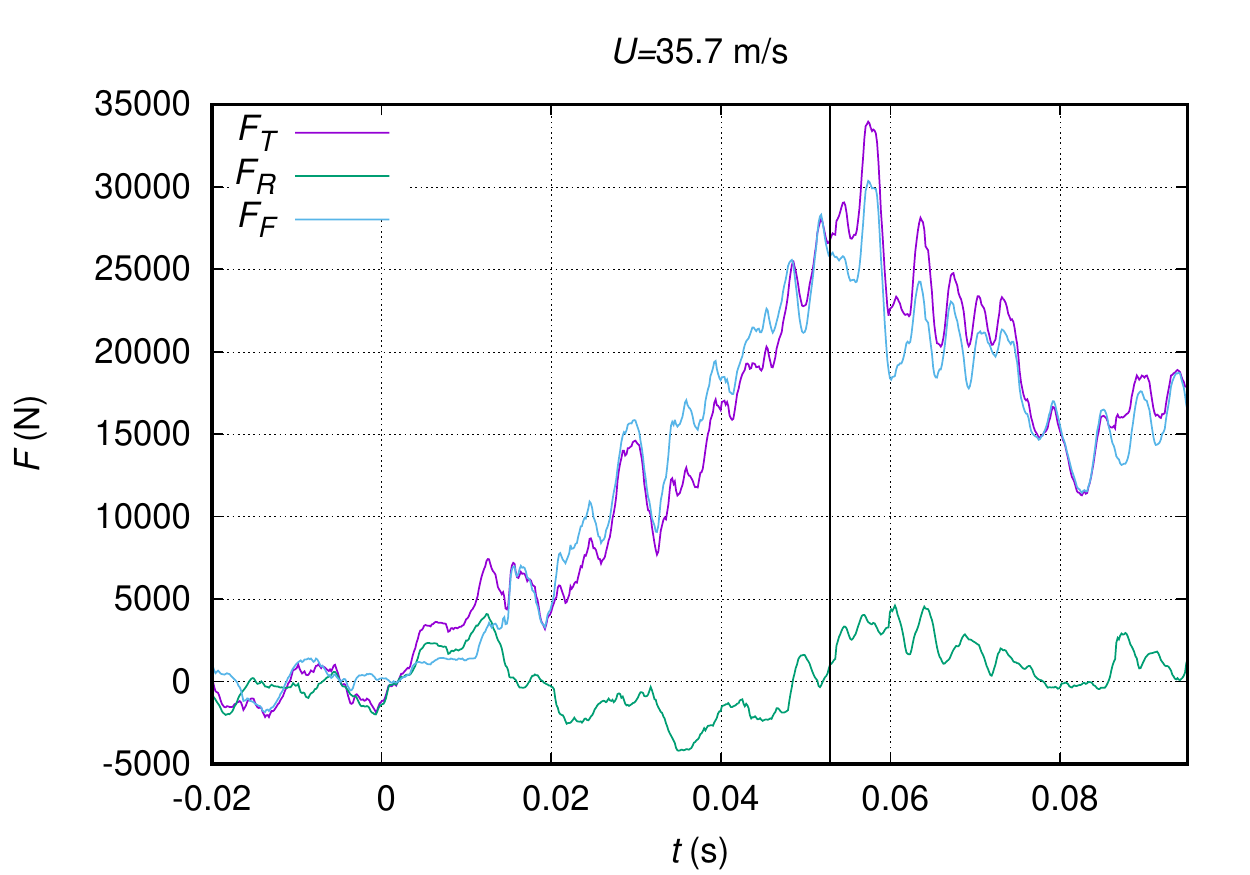}
$f$) \includegraphics[width=80mm]{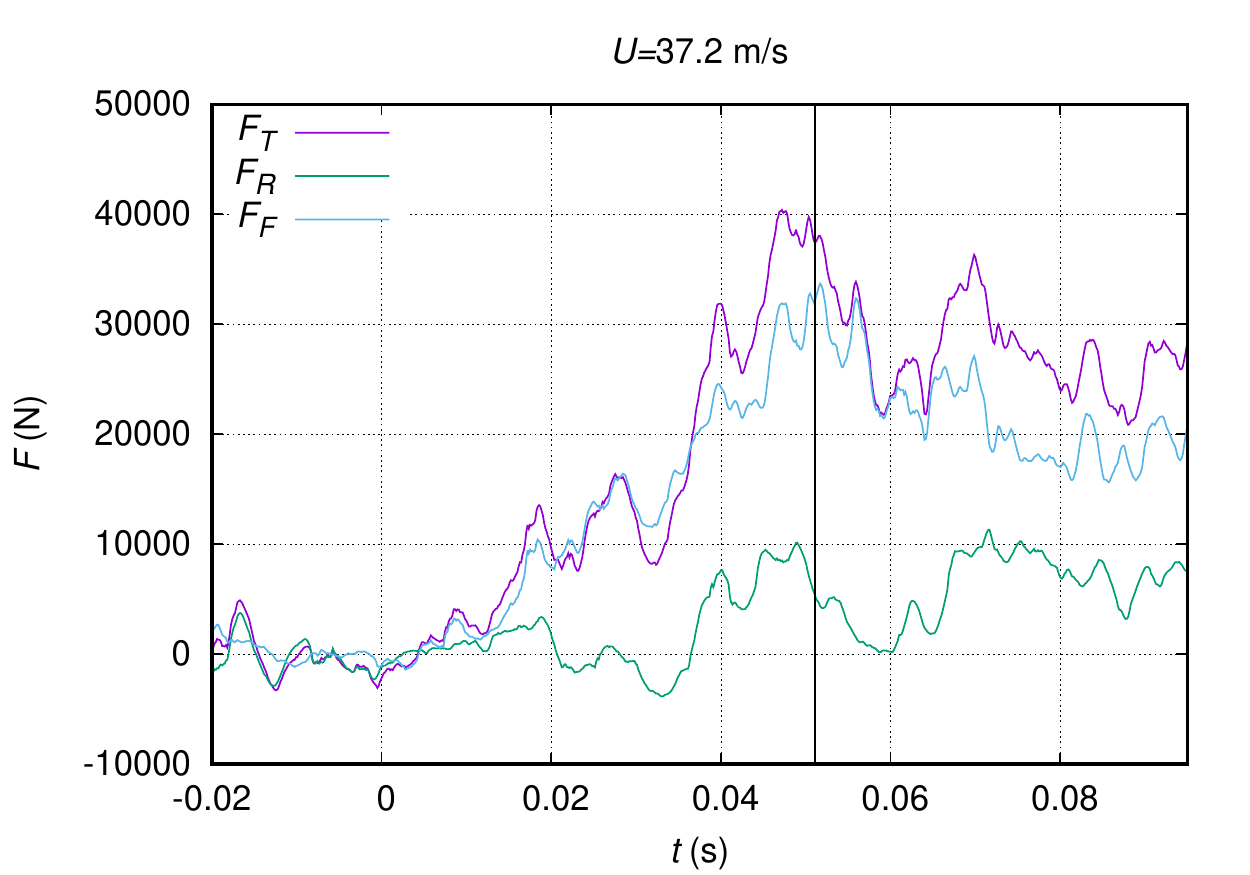}
}
\centerline{
$g$) \includegraphics[width=80mm]{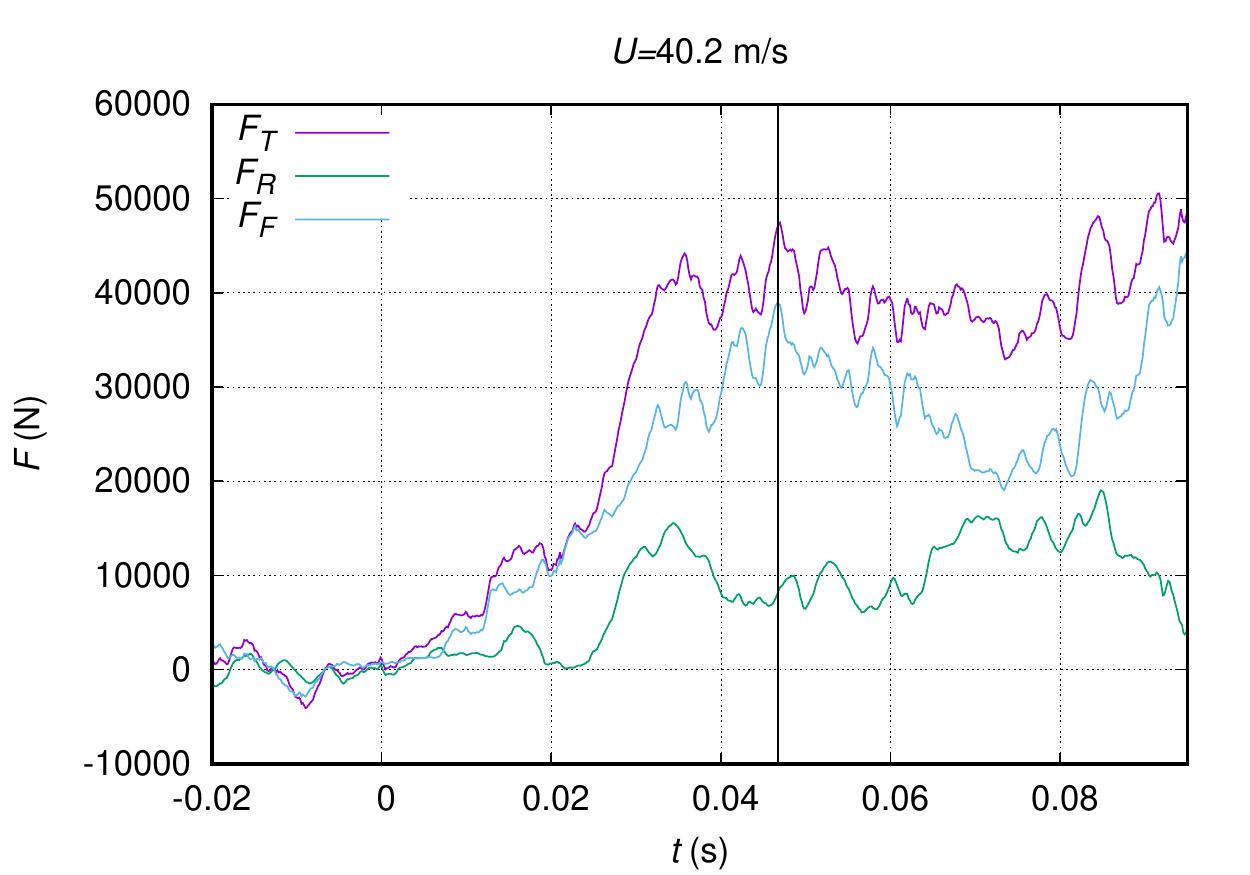}
$h$) \includegraphics[width=80mm]{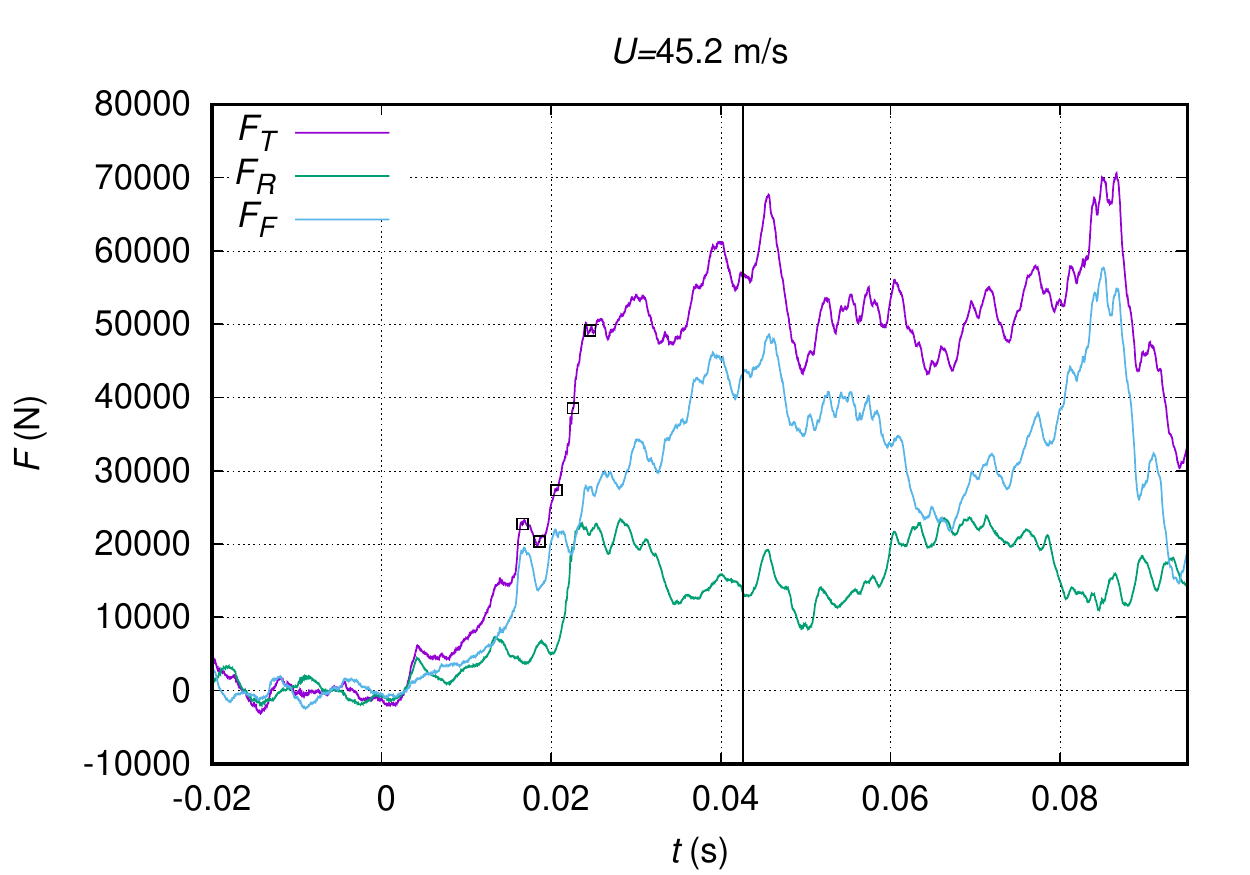}
}
\caption{Time histories of the normal forces acting on the specimen.
On $g)$ the empty boxes drawn on $F_T$ corresponds
to the times of the underwater images shown in figure \ref{fig:cav_ventil}.
}
\label{fig:nor_for_vel}
\end{figure*}

The problem can be examined by analyzing what happens in two-dimensional
planes, and this can be done either in the cross plane or in the
longitudinal plane.
The former is often exploited in the so called 2D+t assumption, like that
used for instance in Ref. \onlinecite{tassin2013, battistin2003}, within 
the potential
flow approximation.
Let us consider the water particles lying on an earth-fixed cross plane
when
the body first touches the water surface, and let us assume that the body
has a double curvature, like that considered in Ref. 
\onlinecite{mcbride1953}, and has a positive pitch angle.
When the body moves forward, the liquid particles in the cross plane are
pushed downward by vertical velocity component of the body $V$, plus the
vertical velocity component $U \tan \alpha$, $\alpha$ being the local slope
of the bottom line in the longitudinal plane.
Owing to the longitudinal profile, the local slope is positive the in the
fore part and turns negative at the rear. Therefore, there is a progressive
transition from a water entry to a water exit phase\cite{tassin2013}.
Because of the upward motion of the cross section of the body, and due to
the inertia of the fluid moving downward, a negative pressure at the rear
occurs, which is a combined effect of the upward velocity and of the upward
acceleration. The latter depends on the product of the curvature of the
bottom profile by the square of the horizontal velocity.

\begin{figure}
\centerline{
\includegraphics[width=40mm,angle=-90]{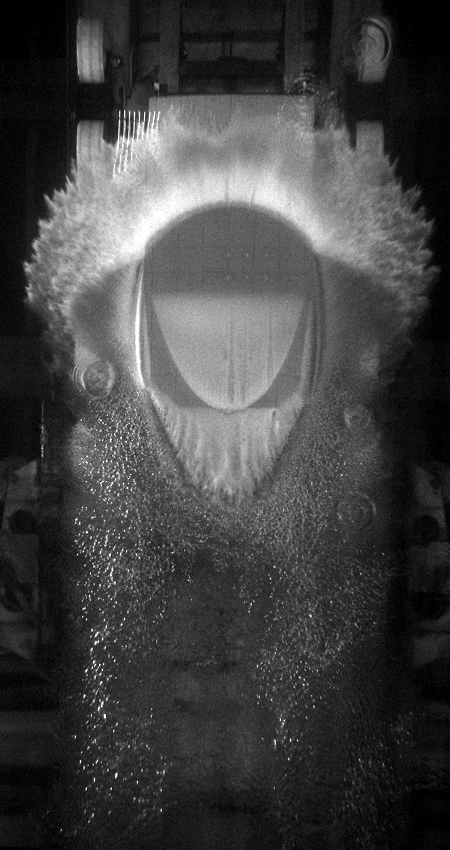}
}
\centerline{
\includegraphics[width=40mm,angle=-90]{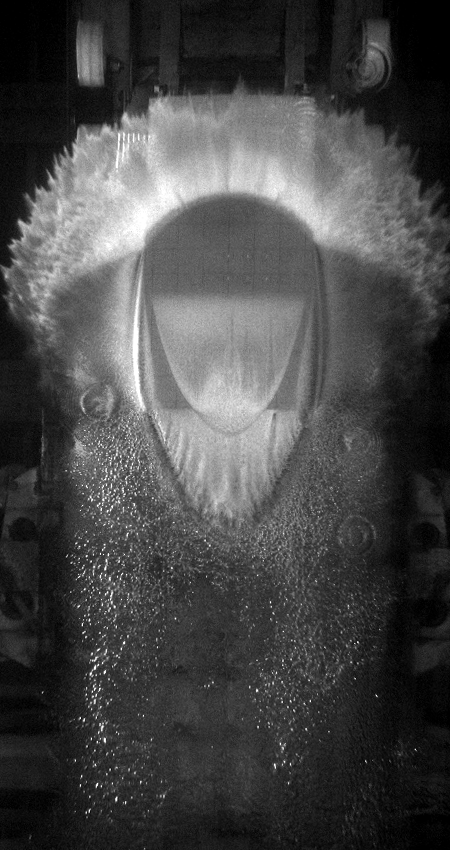}
}
\centerline{
\includegraphics[width=40mm,angle=-90]{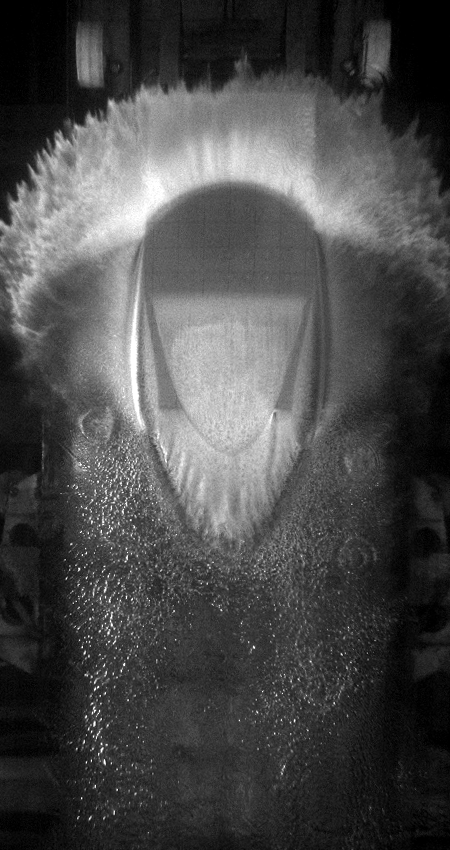}
}
\centerline{
\includegraphics[width=40mm,angle=-90]{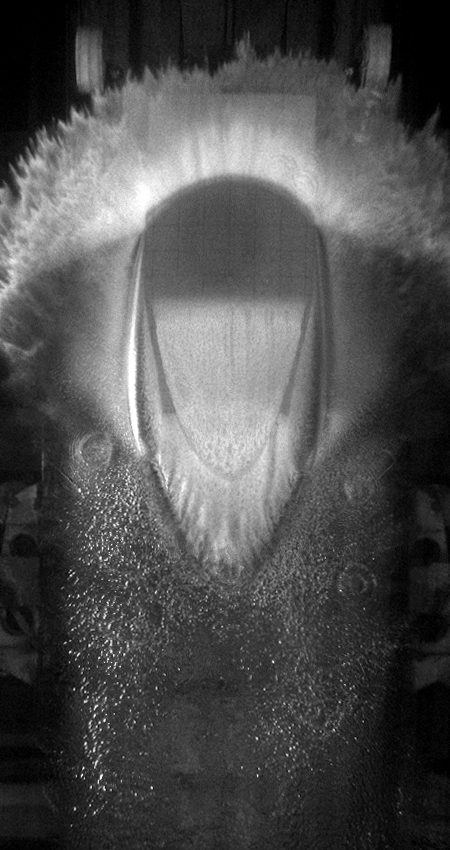}
}
\centerline{
\includegraphics[width=40mm,angle=-90]{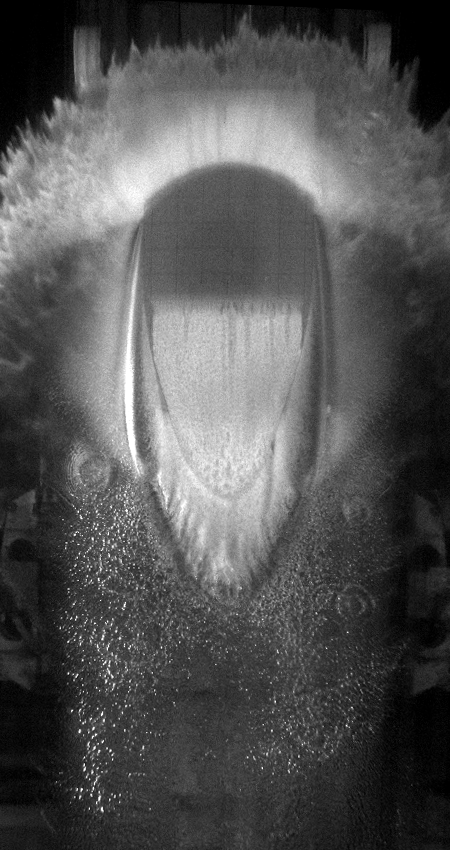}
}
\caption{Sequence of the underwater images showing the formation and the
propagation of the ventilated cavity for the tests at 45 m/s.
The five pictures correspond
to the times indicated in the curves of figures \ref{fig:press_vel}$g$ and
\ref{fig:nor_for_vel}$g$.
}
\label{fig:cav_ventil}
\end{figure}

Although the above representation is rather intuitive and it is indeed the
most used, it is unable to account for the flow taking place in the
longitudinal direction as a consequence of the pressure differences induced
by the asymmetry of the flow \cite{semenov2006}.
As anticipated before, the shape considered in the present study is
characterized by a single curvature portion in the fore part, followed by
a double curvature zone at the rear.
In the longitudinal plane, the fore part can be approximated by a flat
plate, 6 degrees pitch, impacting the water with a velocity ratio
$V/U=0.0375$. The self-similar solution to such problem is derived in
Ref. \onlinecite{iafrati2016} by using a fully nonlinear potential flow 
model and compared to the experimental measurements.
It is shown that the self-similar solution over predicts the propagation
velocity of the spray root along the plate and, consequently, the intensity
of the pressure peak but the pressure profile along the plate is well
captured by the model provided a suitable scaling is adopted.

\begin{figure}
\centerline{
$a)$ \includegraphics[width=70mm]{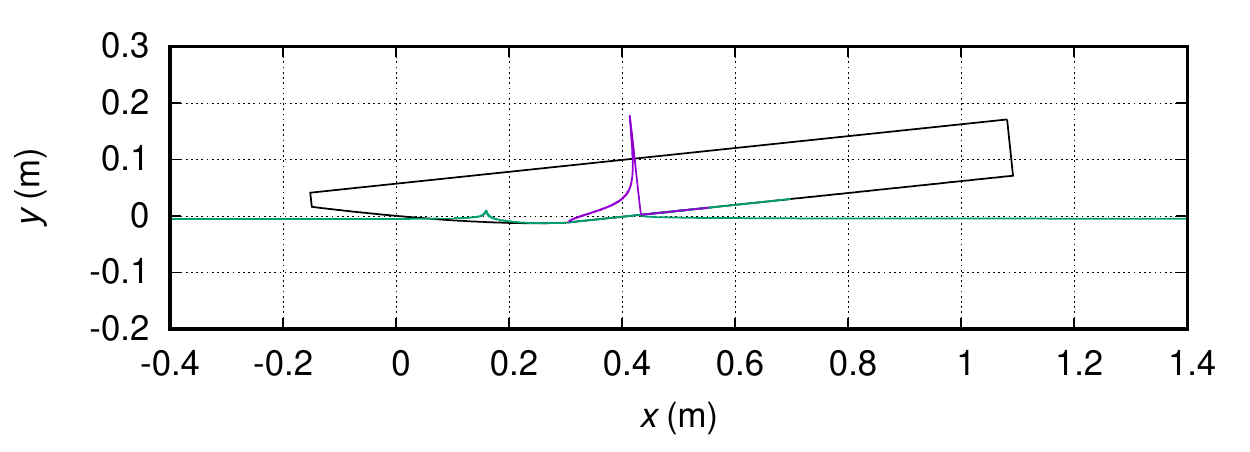}
}
\centerline{
$b)$ \includegraphics[width=70mm]{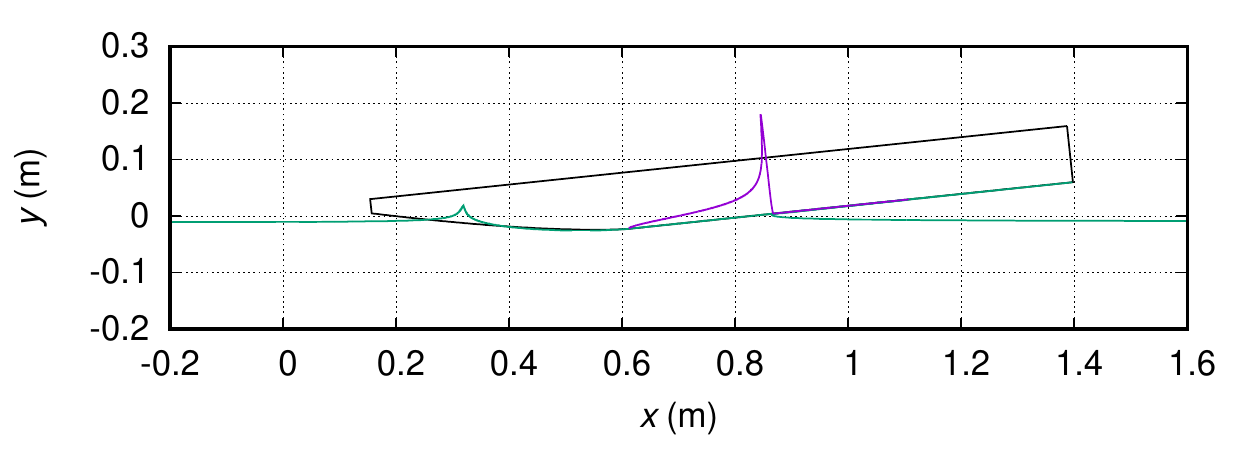}
}
\centerline{
$c)$ \includegraphics[width=70mm]{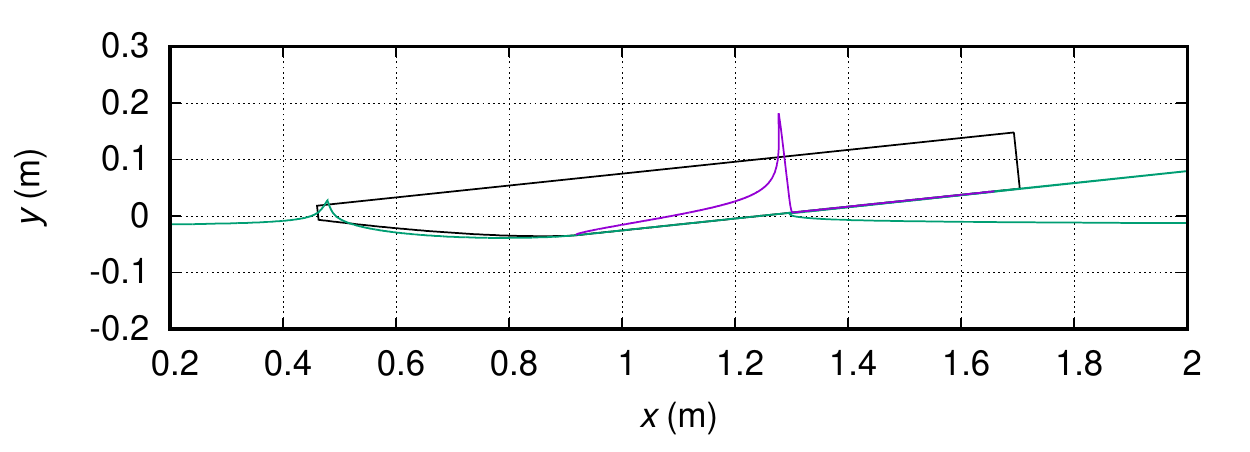}
}
\centerline{
$d)$ \includegraphics[width=70mm]{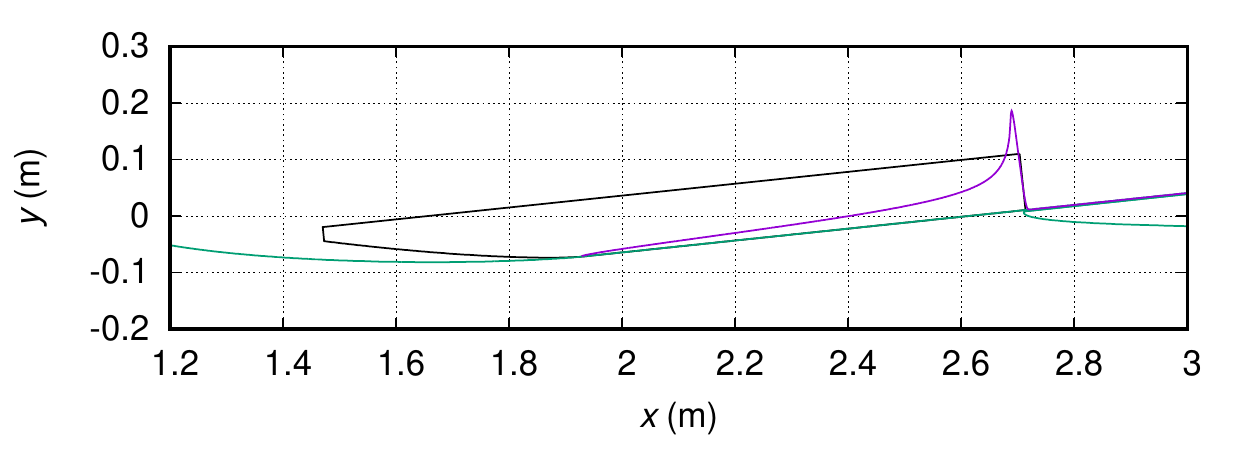}
}
\caption{Self-similar solution of the flat plate problem drawn around the
longitudinal profile of the body at different times for the case at
$U=30.6$ m/s. From top to bottom, time is $10, 20, 30$ and $63$ ms, the
latter denoting the end of the impact phase. The pink line is the
self-similar pressure distribution (not in scale).
}
\label{fig:comp_self_sim}
\end{figure}

Bearing the above considerations in mind, the free-surface shape provided
by the self-similar solution for the flat plate is scaled so that trailing
edge of the plate is located at the end of the single curvature portion,
i.e. at $x=x_H$, whereas the spray root is at the leading edge of the
specimen at the end of the impact phase
(see figure \ref{fig:comp_self_sim}$d$).
The self-similar solution is then further scaled by the corresponding time
fractions and drawn beneath the longitudinal profile of the specimen at
different times for the test at $U=30.6$ m/s
(figure \ref{fig:comp_self_sim}$a,b,c$).
The self-similar solution is drawn in terms of both free-surface shape and
pressure distribution. In the figures, the pressure is not in scale and it
is drawn to give an estimate of the point where the pressure peak is
located.

In figure \ref{fig:comp_self_sim}$a$, which is at $t=0.01$ s from the
initial contact, it is shown that the free surface shed by the trailing
edge
is inside the fuselage profile. This implies that in the impact of the
fuselage specimen, the free surface hits the body surface and thus positive
pressures should be expected, as it happens indeed looking at figure
\ref{fig:press_vel}$c$. The rear bump of the free surface profile is about
90 mm behind the first contact point and that explains why the pressure
at P13 and P9 is relatively high, whereas the pressure at P17 has already
drop, although it is still positive.

Figure \ref{fig:comp_self_sim}$b$, which is at $t=0.02$ s, shows that just
behind the lowest point of the fuselage profile, the free surface has
already
detached from the body and thus negative pressures, and even cavitation
may occur. This is actually confirmed by the time history of the pressure
measured at P17 which drops to the vapor pressure exactly at that time.
However, the free surface profile still intersect the body contour at the
rear, which explains why P4 and P9 still have higher pressures.
At $t=0.03$ s, figure \ref{fig:comp_self_sim}$c$ indicates that the
free surface bump is
at the trailing edge of the specimen and all probes, but for P4, are in
cavitating conditions.
Once the rear bump of the free surface profile leaves the body, the cavity
gets in contact with the atmospheric pressure and the ventilation
start. Of course similar considerations also hold for lower impact
velocities but since the pressure scale with the square of the velocity,
only negative pressures, much above the vapor pressure value, are found.

From the self-similar solution it is seen that the free surface
leaves the trailing edge with a tangential velocity which is about 1/11 of
the horizontal velocity. If the fluid particles have to follow the
curvature
of the body, there should be a normal pressure gradient toward the body
equal to the square of the tangential velocity divided by the local
curvature.
Hence, this is an additional effect that contributes to lower the pressure
together with the upward velocity and accelerations observed in the cross
plane. It is the combination of the different effects in the cross
and in the tangential plane that makes full three-dimensional fully
nonlinear flow solvers essential for a precise picture of the problem.

\section{Conclusions}

The hydrodynamics taking place beneath the aircraft fuselage during the
emergency landing on water has been experimentally investigated. To avoid
scaling issues, realistic shapes and impact velocities have been
considered.
The study has revealed that cavitation and ventilation conditions occur
when the horizontal velocity is above a threshold limit which depends on
the body shape. Different cavitation and ventilation modalities have been
observed, with the transition from one to another taking place within
small velocity variations.

The analysis of pressure and forces acting on the specimen has shown that
the occurrence of cavitation and ventilation significantly affects the
distribution of the loads and thus the aircraft dynamics.
Although the specific results presented depends on the shape and
on the test conditions adopted, e.g. the fixed attitude of the specimen,
the outcomes suggest that an accurate modelling of the hydrodynamics,
either in the experiments or in the numerical simulation,
is essential for a correct prediction of the aircraft behavior at ditching.

\begin{acknowledgments}
This project has received funding from the European Union's
Horizon 2020 Research and Innovation Programme under grant agreement
No 724139 (H2020-SARAH: \emph{increased SAfety \& Robust certification
for ditching of Aircrafts \& Helicopters})
\end{acknowledgments}



\begin{thebibliography}{26}
\expandafter\ifx\csname natexlab\endcsname\relax\def\natexlab#1{#1}\fi
\expandafter\ifx\csname bibnamefont\endcsname\relax
  \def\bibnamefont#1{#1}\fi
\expandafter\ifx\csname bibfnamefont\endcsname\relax
  \def\bibfnamefont#1{#1}\fi
\expandafter\ifx\csname citenamefont\endcsname\relax
  \def\citenamefont#1{#1}\fi
\expandafter\ifx\csname url\endcsname\relax
  \def\url#1{\texttt{#1}}\fi
\expandafter\ifx\csname urlprefix\endcsname\relax\def\urlprefix{URL }\fi
\providecommand{\bibinfo}[2]{#2}
\providecommand{\eprint}[2][]{\url{#2}}

\bibitem[{\citenamefont{Siemann and Langrand}(2017)}]{siemann2017}
\bibinfo{author}{\bibfnamefont{M.}~\bibnamefont{Siemann}} \bibnamefont{and}
  \bibinfo{author}{\bibfnamefont{B.}~\bibnamefont{Langrand}}, Coupled
  fluid-structure computational methods for aircraft ditching simulations:
  Comparison of ale-fe and sph-fe approaches, \bibinfo{journal}{Comput Struct}
  \textbf{\bibinfo{volume}{188}}, \bibinfo{pages}{95} (\bibinfo{year}{2017}).

\bibitem[{\citenamefont{Siemann et~al.}(2018)\citenamefont{Siemann, Schwinn,
  Scherer, and Kohlgr{\"u}ber}}]{siemann2018}
\bibinfo{author}{\bibfnamefont{M.}~\bibnamefont{Siemann}},
  \bibinfo{author}{\bibfnamefont{D.}~\bibnamefont{Schwinn}},
  \bibinfo{author}{\bibfnamefont{J.}~\bibnamefont{Scherer}}, \bibnamefont{and}
  \bibinfo{author}{\bibfnamefont{D.}~\bibnamefont{Kohlgr{\"u}ber}}, Advances in
  numerical ditching simulation of flexible aircraft models,
  \bibinfo{journal}{Int J Crashworthines} \textbf{\bibinfo{volume}{23}},
  \bibinfo{pages}{236} (\bibinfo{year}{2018}).

\bibitem[{\citenamefont{Qu et~al.}(2015)\citenamefont{Qu, Hu, Guo, Liu, and
  Agarwal}}]{qu2015}
\bibinfo{author}{\bibfnamefont{Q.}~\bibnamefont{Qu}},
  \bibinfo{author}{\bibfnamefont{M.}~\bibnamefont{Hu}},
  \bibinfo{author}{\bibfnamefont{H.}~\bibnamefont{Guo}},
  \bibinfo{author}{\bibfnamefont{P.}~\bibnamefont{Liu}}, \bibnamefont{and}
  \bibinfo{author}{\bibfnamefont{R.}~\bibnamefont{Agarwal}}, Study of ditching
  characteristics of transport aircraft by global moving mesh method,
  \bibinfo{journal}{J Aircraft} \textbf{\bibinfo{volume}{52}},
  \bibinfo{pages}{1550} (\bibinfo{year}{2015}).

\bibitem[{\citenamefont{Bisagni and Pigazzini}(2018)}]{bisagni2018}
\bibinfo{author}{\bibfnamefont{C.}~\bibnamefont{Bisagni}} \bibnamefont{and}
  \bibinfo{author}{\bibfnamefont{M.}~\bibnamefont{Pigazzini}}, Modelling
  strategies for numerical simulation of aircraft ditching,
  \bibinfo{journal}{Int J Crashworthines} \textbf{\bibinfo{volume}{23}},
  \bibinfo{pages}{377} (\bibinfo{year}{2018}).

\bibitem[{\citenamefont{Landau and Lifshitz}(1959)}]{landau1959}
\bibinfo{author}{\bibfnamefont{L.~D.} \bibnamefont{Landau}} \bibnamefont{and}
  \bibinfo{author}{\bibfnamefont{E.~M.} \bibnamefont{Lifshitz}},
  \emph{\bibinfo{title}{Fluid mechanics}}, \bibinfo{howpublished}{Pergamon},
  \bibinfo{address}{New York} (\bibinfo{year}{1959}).

\bibitem[{\citenamefont{Moghisi and Squire}(1981)}]{Moghisi1981}
\bibinfo{author}{\bibfnamefont{M.}~\bibnamefont{Moghisi}} \bibnamefont{and}
  \bibinfo{author}{\bibfnamefont{P.}~\bibnamefont{Squire}}, An experimental
  investigation of the initial force of impact on a sphere striking a liquid
  surface, \bibinfo{journal}{J Fluid Mech} \textbf{\bibinfo{volume}{108}},
  \bibinfo{pages}{133} (\bibinfo{year}{1981}).

\bibitem[{\citenamefont{Smith et~al.}(1957)\citenamefont{Smith, Warren, and
  Wright}}]{smith1957}
\bibinfo{author}{\bibfnamefont{A.~G.} \bibnamefont{Smith}},
  \bibinfo{author}{\bibfnamefont{C.~H.~E.} \bibnamefont{Warren}},
  \bibnamefont{and} \bibinfo{author}{\bibfnamefont{D.~F.}
  \bibnamefont{Wright}}, \bibinfo{type}{Reports and Memoranda}
  \bibinfo{number}{2917}, \bibinfo{institution}{Aeronautical Research Council},
  \bibinfo{address}{Ministry of Supply} (\bibinfo{year}{1957}).

\bibitem[{\citenamefont{Zhang et~al.}(2012)\citenamefont{Zhang, Li, and
  Dai}}]{zhang2012}
\bibinfo{author}{\bibfnamefont{T.}~\bibnamefont{Zhang}},
  \bibinfo{author}{\bibfnamefont{S.}~\bibnamefont{Li}}, \bibnamefont{and}
  \bibinfo{author}{\bibfnamefont{H.}~\bibnamefont{Dai}}, The suction force
  effect analysis of large civil aircraft ditching, \bibinfo{journal}{Science
  China - Technological Sciences} \textbf{\bibinfo{volume}{55}},
  \bibinfo{pages}{2789} (\bibinfo{year}{2012}).

\bibitem[{\citenamefont{Climent et~al.}(2006)\citenamefont{Climent, Benitez,
  Rosich, Rueda, and Pentecote}}]{climent2006}
\bibinfo{author}{\bibfnamefont{H.}~\bibnamefont{Climent}},
  \bibinfo{author}{\bibfnamefont{L.}~\bibnamefont{Benitez}},
  \bibinfo{author}{\bibfnamefont{F.}~\bibnamefont{Rosich}},
  \bibinfo{author}{\bibfnamefont{F.}~\bibnamefont{Rueda}}, \bibnamefont{and}
  \bibinfo{author}{\bibfnamefont{N.}~\bibnamefont{Pentecote}}, in
  \emph{\bibinfo{booktitle}{Proceedings of the 25$^{th}$ International Congress
  of the Aeronautical Sciences}} (\bibinfo{address}{Hamburg, Germany},
  \bibinfo{year}{2006}).

\bibitem[{\citenamefont{Iafrati et~al.}(2015)\citenamefont{Iafrati, Grizzi,
  Siemann, and {Ben\'itez~Monta\~n\'es}}}]{iafratijfs2015}
\bibinfo{author}{\bibfnamefont{A.}~\bibnamefont{Iafrati}},
  \bibinfo{author}{\bibfnamefont{S.}~\bibnamefont{Grizzi}},
  \bibinfo{author}{\bibfnamefont{M.}~\bibnamefont{Siemann}}, \bibnamefont{and}
  \bibinfo{author}{\bibfnamefont{L.}~\bibnamefont{{Ben\'itez~Monta\~n\'es}}},
  High-speed ditching of a flat plate: Experimental data and uncertainty
  assessment, \bibinfo{journal}{J Fluid Struct} \textbf{\bibinfo{volume}{55}},
  \bibinfo{pages}{501} (\bibinfo{year}{2015}).

\bibitem[{\citenamefont{Iafrati}(2016{\natexlab{a}})}]{iafratijfm2016}
\bibinfo{author}{\bibfnamefont{A.}~\bibnamefont{Iafrati}}, Experimental
  investigation of the water entry of a rectangular plate at high horizontal
  velocity, \bibinfo{journal}{J Fluid Mech} \textbf{\bibinfo{volume}{799}},
  \bibinfo{pages}{637} (\bibinfo{year}{2016}{\natexlab{a}}).

\bibitem[{\citenamefont{Iafrati}(2015{\natexlab{a}})}]{iafratinav2015}
\bibinfo{author}{\bibfnamefont{A.}~\bibnamefont{Iafrati}}, in
  \emph{\bibinfo{booktitle}{Proceedings of the 18$^{th}$ International
  Conference on Ships and Shipping Research}}
  (\bibinfo{year}{2015}{\natexlab{a}}).

\bibitem[{\citenamefont{Iafrati}(2018)}]{iafrati2018}
\bibinfo{author}{\bibfnamefont{A.}~\bibnamefont{Iafrati}}, in
  \emph{\bibinfo{booktitle}{Proceedings of the ASME 2018 37th International
  Conference on Ocean, Offshore and Arctic Engineering}}
  (\bibinfo{address}{Madrid, Spain}, \bibinfo{year}{2018}),
  \bibinfo{number}{OMAE2018-78438}.

\bibitem[{\citenamefont{McBride and Fisher}(1953)}]{mcbride1953}
\bibinfo{author}{\bibfnamefont{E.}~\bibnamefont{McBride}} \bibnamefont{and}
  \bibinfo{author}{\bibfnamefont{L.}~\bibnamefont{Fisher}},
  \bibinfo{type}{Technical Note} \bibinfo{number}{2929},
  \bibinfo{institution}{National Advisory Council for Aeronautics (NACA)},
  \bibinfo{address}{Langley Field, VA, USA} (\bibinfo{year}{1953}).

\bibitem[{\citenamefont{Tassin et~al.}(2013)\citenamefont{Tassin, Piro,
  Korobkin, Maki, and Cooker}}]{tassin2013}
\bibinfo{author}{\bibfnamefont{A.}~\bibnamefont{Tassin}},
  \bibinfo{author}{\bibfnamefont{D.}~\bibnamefont{Piro}},
  \bibinfo{author}{\bibfnamefont{A.}~\bibnamefont{Korobkin}},
  \bibinfo{author}{\bibfnamefont{K.}~\bibnamefont{Maki}}, \bibnamefont{and}
  \bibinfo{author}{\bibfnamefont{M.}~\bibnamefont{Cooker}}, Two-dimensional
  water entry and exit of a body whose shape varies in time,
  \bibinfo{journal}{J Fluid Struct} \textbf{\bibinfo{volume}{40}},
  \bibinfo{pages}{317} (\bibinfo{year}{2013}).

\bibitem[{\citenamefont{Wang et~al.}(2010)\citenamefont{Wang, Xu, Wu, Huang, ,
  and Wu}}]{wang2017}
\bibinfo{author}{\bibfnamefont{Y.}~\bibnamefont{Wang}},
  \bibinfo{author}{\bibfnamefont{C.}~\bibnamefont{Xu}},
  \bibinfo{author}{\bibfnamefont{X.}~\bibnamefont{Wu}},
  \bibinfo{author}{\bibfnamefont{C.}~\bibnamefont{Huang}}, , \bibnamefont{and}
  \bibinfo{author}{\bibfnamefont{X.}~\bibnamefont{Wu}}, Ventilated cloud
  cavitating flow around a blunt body close to the free surface,
  \bibinfo{journal}{Phys Rev Fluids} \textbf{\bibinfo{volume}{2}},
  \bibinfo{pages}{084303} (\bibinfo{year}{2010}).

\bibitem[{\citenamefont{Ceccio}(2010)}]{ceccio2010}
\bibinfo{author}{\bibfnamefont{S.}~\bibnamefont{Ceccio}}, Friction drag
  reduction of external flows with bubble and gas injection,
  \bibinfo{journal}{Annu Rev Fluid Mech} \textbf{\bibinfo{volume}{42}},
  \bibinfo{pages}{183} (\bibinfo{year}{2010}).

\bibitem[{\citenamefont{Iafrati and Olivieri}(2017)}]{iafrati2017}
\bibinfo{author}{\bibfnamefont{A.}~\bibnamefont{Iafrati}} \bibnamefont{and}
  \bibinfo{author}{\bibfnamefont{F.}~\bibnamefont{Olivieri}},
  \bibinfo{type}{SARAH Deliverable D5.1}, \bibinfo{institution}{Institute of
  Marine Engineering (CNR-INM)}, \bibinfo{address}{Rome, Italy}
  (\bibinfo{year}{2017}).

\bibitem[{\citenamefont{Iafrati et~al.}(2019)\citenamefont{Iafrati, Grizzi, and
  Olivieri}}]{iafratiaiaa}
\bibinfo{author}{\bibfnamefont{A.}~\bibnamefont{Iafrati}},
  \bibinfo{author}{\bibfnamefont{S.}~\bibnamefont{Grizzi}}, \bibnamefont{and}
  \bibinfo{author}{\bibfnamefont{F.}~\bibnamefont{Olivieri}}, in
  \emph{\bibinfo{booktitle}{Proceedings of the 2019 SciTech Forum}}
  (\bibinfo{address}{San Diego (CA), USA}, \bibinfo{year}{2019}),
  \bibinfo{number}{AIAA-2019-2030}.

\bibitem[{\citenamefont{{Van Nuffel} et~al.}(2013)\citenamefont{{Van Nuffel},
  Vepa, {De Baere}, Degrieck, {De Rouck}, and {Van Paepegem}}}]{vanhuffel2013}
\bibinfo{author}{\bibfnamefont{D.}~\bibnamefont{{Van Nuffel}}},
  \bibinfo{author}{\bibfnamefont{K.~S.} \bibnamefont{Vepa}},
  \bibinfo{author}{\bibfnamefont{I.}~\bibnamefont{{De Baere}}},
  \bibinfo{author}{\bibfnamefont{J.}~\bibnamefont{Degrieck}},
  \bibinfo{author}{\bibfnamefont{J.}~\bibnamefont{{De Rouck}}},
  \bibnamefont{and} \bibinfo{author}{\bibfnamefont{W.}~\bibnamefont{{Van
  Paepegem}}}, Study on the parameters influencing the accuracy and
  reproducibility of dynamic pressure measurements at the surface of a rigid
  body during water impact, \bibinfo{journal}{Exp. Mech.}
  \textbf{\bibinfo{volume}{53}}, \bibinfo{pages}{131} (\bibinfo{year}{2013}).

\bibitem[{\citenamefont{Iafrati}(2016{\natexlab{b}})}]{iafrati2016}
\bibinfo{author}{\bibfnamefont{A.}~\bibnamefont{Iafrati}}, Experimental
  investigation of the water entry of a rectangular plate at high horizontal
  velocity, \bibinfo{journal}{J Fluid Mech} \textbf{\bibinfo{volume}{799}},
  \bibinfo{pages}{637} (\bibinfo{year}{2016}{\natexlab{b}}).

\bibitem[{\citenamefont{Semenov and Iafrati}(2006)}]{semenov2006}
\bibinfo{author}{\bibfnamefont{Y.~A.} \bibnamefont{Semenov}} \bibnamefont{and}
  \bibinfo{author}{\bibfnamefont{A.}~\bibnamefont{Iafrati}}, On the nonlinear
  water entry problem of asymmetric wedges, \bibinfo{journal}{J Fluid Mech}
  \textbf{\bibinfo{volume}{547}}, \bibinfo{pages}{231} (\bibinfo{year}{2006}).

\bibitem[{\citenamefont{Riccardi and Iafrati}(2004)}]{riccardi2004}
\bibinfo{author}{\bibfnamefont{G.}~\bibnamefont{Riccardi}} \bibnamefont{and}
  \bibinfo{author}{\bibfnamefont{A.}~\bibnamefont{Iafrati}}, Water impact of an
  asymmetric floating wedge, \bibinfo{journal}{J Eng Math}
  \textbf{\bibinfo{volume}{49}}, \bibinfo{pages}{19} (\bibinfo{year}{2004}).

\bibitem[{\citenamefont{Judge et~al.}(2004)\citenamefont{Judge, Troesch, and
  Perlin}}]{judge2004}
\bibinfo{author}{\bibfnamefont{C.}~\bibnamefont{Judge}},
  \bibinfo{author}{\bibfnamefont{A.}~\bibnamefont{Troesch}}, \bibnamefont{and}
  \bibinfo{author}{\bibfnamefont{M.}~\bibnamefont{Perlin}}, Initial water
  impact of a wedge at vertical and oblique angles, \bibinfo{journal}{J Eng
  Math} \textbf{\bibinfo{volume}{48}}, \bibinfo{pages}{279}
  (\bibinfo{year}{2004}).

\bibitem[{\citenamefont{Iafrati}(2015{\natexlab{b}})}]{iafrati2015}
\bibinfo{author}{\bibfnamefont{A.}~\bibnamefont{Iafrati}}, in
  \emph{\bibinfo{booktitle}{Proceedings of the 18th International Conference on
  Ships and Shipping Research}} (\bibinfo{address}{Lecco, Italy},
  \bibinfo{year}{2015}{\natexlab{b}}).

\bibitem[{\citenamefont{Battistin and Iafrati}(2003)}]{battistin2003}
\bibinfo{author}{\bibfnamefont{D.}~\bibnamefont{Battistin}} \bibnamefont{and}
  \bibinfo{author}{\bibfnamefont{A.}~\bibnamefont{Iafrati}}, Hydrodynamic loads
  during water entry of two-dimensional and axisymmetric bodies,
  \bibinfo{journal}{J Fluid Struct} \textbf{\bibinfo{volume}{17}},
  \bibinfo{pages}{643} (\bibinfo{year}{2003}).

\end{thebibliography}
\end{document}